\documentclass[aps,twocolumn,pra,amsmath,amssymb,superscriptaddress,floats,nobibnotes,nofootinbib]{revtex4}

\usepackage{CJK}
\usepackage{amsmath}	
\usepackage{gensymb}
\usepackage{hyperref}
\hypersetup{colorlinks=true}
\usepackage{upgreek}
\usepackage{graphics}
\usepackage{hyperref}
\usepackage{epsfig}
\usepackage{color}
\usepackage{bm}
\usepackage{float}
\usepackage{ulem} 
\usepackage{dsfont}                 
\usepackage{wasysym}
\usepackage{multirow}

\usepackage{blindtext}

\usepackage{accents}

\usepackage{graphicx}

\graphicspath{{./}{./plots/}}

\usepackage{url}
\usepackage{multirow}

\def\infinity{\infty}

\newcommand{\upperRomannumeral}[1]{\uppercase\expandafter{\romannumeral#1}}

\begin{document}
\title{Topological phase transitions and thermal Hall effect in a noncollinear spin texture}

\author{Ken Chen}
\affiliation{School of Physical Science and Technology $\&$ Key Laboratory of Quantum Theory and Applications of MoE, Lanzhou University, Lanzhou 730000, China}
\affiliation{Lanzhou Center for Theoretical Physics, Key Laboratory of Theoretical Physics of Gansu Province, Lanzhou University, Lanzhou, Gansu 730000, China}

\author{Qiang Luo}
\email[]{qiangluo@nuaa.edu.cn}
\affiliation{College of Physics, Nanjing University of Aeronautics and Astronautis, Nanjing, 211106, China}

\author{Bin Xi}
\affiliation{College of Physics Science and Technology, Yangzhou University, Yangzhou 225002, China}

\author{Hong-Gang Luo}
\affiliation{School of Physical Science and Technology $\&$ Key Laboratory of Quantum Theory and Applications of MoE, Lanzhou University, Lanzhou 730000, China}
\affiliation{Lanzhou Center for Theoretical Physics, Key Laboratory of Theoretical Physics of Gansu Province, Lanzhou University, Lanzhou, Gansu 730000, China}

\author{Jize Zhao}
\email[]{zhaojz@lzu.edu.cn}
\affiliation{School of Physical Science and Technology $\&$ Key Laboratory of Quantum Theory and Applications of MoE, Lanzhou University, Lanzhou 730000, China}
\affiliation{Lanzhou Center for Theoretical Physics, Key Laboratory of Theoretical Physics of Gansu Province, Lanzhou University, Lanzhou, Gansu 730000, China}

\date{\today}
\begin{abstract}
  The noncollinear spin textures provide promising avenues to stabilize exotic magnetic phases and excitations.
  They have attracted vast attention owning to their nontrivial band topology in the past decades.
  Distinct from the conventional route of involving the Dzyaloshinskii-Moriya interaction in a honeycomb magnet,
  the interplay of bond-dependent Kitaev and $\Gamma$ interactions,
  originating from the spin-orbit coupling and octahedra crystal field in real materials,
  has demonstrated to be another source to generate noncollinear spin textures with multiple spins in a magnetic unit cell.
  Notably, earlier works have revealed a triple-meron crystal (TmX) consisting of eighteen spins in the frustrated Kitaev-$\Gamma$ model.
  Aligning with previous efforts, here we attempt to identify that the TmX hosts several peculiar features with the help of the linear spin-wave theory.
  To begin with, the symmetric anisotropic exchanges are beneficial for the existence of nonreciprocal magnons, which are stabilized by an external magnetic field.
  Further, within the regime of TmX, successive topological phase transitions occur, accompanied by the changes of Chern number in value and thermal Hall conductivity in sign.
  In addition, topological nature of magnons is also verified by the onset of chiral edge modes in a nanoribbon geometry.
  Our findings pave the way to study topological phenomena of noncollinear spin textures in potential Kitaev materials.
\end{abstract}

\maketitle

\section{Introduction}
The theory of topological band structures has been extended beyond the electronic system
to embrace topological magnon insulators and magnonic Dirac and Weyl semimetals \cite{McClarty2022,PengYan2021,Fengjun2023}.
The magnons are the quanta of the low-energy collective excitations which are ubiquitous in magnetic materials.
They are able to transfer spins without producing Joule heating and are believed to have significant impacts on spintronics serving as ingredients to low-energy consumption devices \cite{Lenk2011,Chumak2015}.
As the inversion or time-reversal symmetry breaks, it is natural to expect the magnon band structure to display nontrivial topological signatures \cite{Lifa2013,Shindou2013,Mook2014,Nakata2017}.
Of note is that a temperature gradient can induce a magnon flow, leading to the thermal Hall effect due to a transversal magnon current through the nonzero Berry curvature \cite{Katsura2010,Qin2011,Matsumoto2011prl,Ideue2012,Matsumoto2014,Mook2014b,Murakami2017,Mook2019,Yang2020,Wen2020,Mook2022,Yao2023,Chen2023}.

The Dzyaloshinskii-Moriya interaction has been well recognized to obtain nontrivial magnon bands \cite{McClarty2022,PengYan2021,Fengjun2023,Lifa2013,Mook2014}.
It not only acts as a virtual magnetic field but introduces an effective non-Abelian gauge field for magnons,
leaving the possibility of nontrivial Berry curvature.
Experimentally, the thermal Hall effect has been observed in various ferromagnetic insulators where the Dzyaloshinskii-Moriya interaction is demonstrated to play a vital role \cite{Onose2010,Chisnell2015,Hirschberger2015,Hirschberger2015prl}.
However, the Dzyaloshinskii-Moriya interaction is an antisymmetric exchange interaction that is induced by inversion symmetry breaking.
It is thus either symmetry-forbidden or usually acquires a small intensity in further nearest-neighbor interactions.
On the other hand, quantum materials with bond-dependent Kitaev-type interactions emerge as the focus of experimental and theoretical studies over the past years \cite{Trebst2022,Winter2017,Wen2017,Janssen2019,XuFXetal2018,LeeUWetal2020,GLin2021,Li2022,Yao2023prr,Jie2023}.
These competing exchange couplings strongly promote the frustration, giving rise to exotic phases of matter
such as quantum spin liquids \cite{WangNmdLiu2019,RalkoMerino2020,LuoNPJ2021} and nematic paramagnet \cite{LeeKCetal2020,GohlkeCKK2020}.
Notably, the Kitaev-type interactions, associated with the spin-orbit coupling, have been interpreted as another source to generate topological magnon excitations \cite{McClarty2018,Joshi2018,Luo2020SciPost,Chern2020,Aguilera2020,Emily2020}.
The magnon bands of Kitaev honeyomb magnets can carry nonzero Chern number and chiral edge modes at high magnetic field \cite{McClarty2018,Joshi2018,Luo2020SciPost}.
In addition, the thermal hall conductivity undergoes a sign change as the direction of the in-plane magnetic field reverses \cite{Chern2021prl}.
Moreover, the abnormal phenomena in a couple of thermal Hall measurements on $\alpha$-RuCl$_3$ \cite{Kasahara2018Nature,Czajka2021NatPhys,Kee2023NatMater},
together with other Kitaev materials like Na$_2$Co$_2$TeO$_6$ \cite{Takeda2022PRR,Guang2023PRB} and MnPS$_3$ \cite{Mook2022},
render the topological magnon as a promising carrier to dominate these tempting behaviors at low temperatures.

Nevertheless, the topological magnon on a honeycomb lattice has been so far mainly studied in strong or modest magnetic field,
at which the underlying spins are parallel or nearly parallel \cite{McClarty2018,Joshi2018,Luo2020SciPost,Chern2021prl,Luo2022PRB}.
The topological magnon in noncollinear spin textures with a large magnetic unit cell at a weak magnetic field thus calls for an urgent study.
It is revealed that the competition between the Kitaev and $\Gamma$ interactions can generate many noncoplanar magnetic orders \cite{Rau2014,Janssen2019,Chern2020,Chern2021npj,Liu2021PRR,Rao2021PRR,Rayyan2022,Ken2023,Stavropoulos2023arXiv,Rousochatzakis2023},
such as the (6+18) state \cite{Chern2020,Rayyan2022,Ken2023}, the nested zigzag-stripy order \cite{Rao2021PRR},
and the $C_3$-like \cite{Rayyan2022} triple-meron crystal (TmX) \cite{Ken2023}.
Among them, the TmX is extremely alluring in that it has three merons within one magnetic unit cell
and it occupies a large area in the phase diagram of the $K$-$\Gamma$ model where $K < 0$ and $\Gamma > 0$ \cite{Rayyan2022,Ken2023}.
Thus, in this work, we focus on the magnon excitation in such order with the help of the linear spin-wave theory.
Our results manifest that nontrivial magnon band topology is widely present within the parameter range of interest.
The competition between Kitaev and $\Gamma$ interactions also produces topological phase transitions within the magnetically ordered phase.
In this respect, the non-reciprocity of magnons is revealed and multiple topological phases are distinguished by the Chern number.
Moreover, we calculate the experimentally observable thermal Hall conductivity and discuss its consistency with band topology at low temperatures.

\section{Model and methods}
\label{Sec:Model}
\begin{figure*}[htb]
    \centering
	\includegraphics[width=1\textwidth]{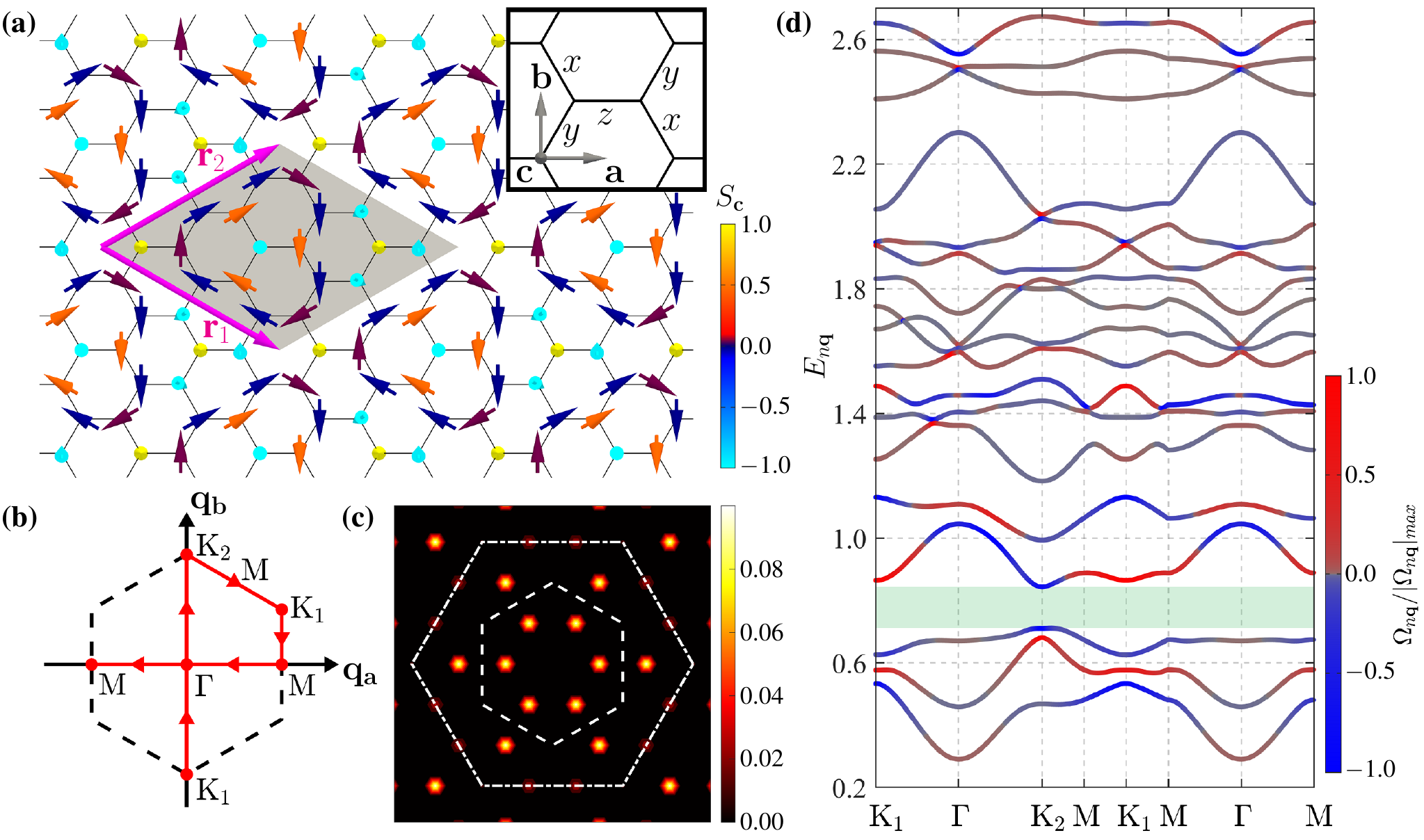}\\
	\caption{(a) Top view of spin configuration in the TmX phase where $\phi = 0.64\pi$ and $h/\left(\mathcal{E}_0 S\right) = 0.1$.
		The small arrows indicate directions of spins and the colors are based on their out of plane components $S_{\mathbf{c}}$.
		The magnetic unit cell is shown in the gray area which includes eighteen spins. Two long pink arrows $\textbf{r}_1$ and $\textbf{r}_2$ represent the primitive vectors.
		The inset indicates the $\{\textbf{a}, \textbf{b}, \textbf{c}\}$ coordinate system
        and three kinds of bonds in the honeycomb lattice are labeled as $x$, $y$, and $z$, respectively.
		(b) The first Brillouin zone is marked by the dashed line. The high-symmetry points and a special path in the reciprocal space are shown.
		(c) The static structure factor of the configurations in (a). The ordering wave vector locates at $2\mathbf{M}/{3}$ point.
		(d) The magnon band structure along the special path in (b) and the color in each band stands for the normalized Berry curvature.
		The green zone declares that the lowest three magnon bands is well separated from the others with a global band gap.
	}\label{FIG-TmXPattern}
\end{figure*}

For the study of the topological magnon in a honeycomb lattice, the model is given by
\begin{equation}\label{Eq:HamKGH}
\mathcal{H} = \sum_{\langle i, j \rangle_{\gamma}} \left[K S^\gamma_i S^\gamma_j + \Gamma(S^\alpha_i S^\beta_j +S^\beta_i S^\alpha_j)\right] + \sum_i \textbf{h}\cdot\mathbf{S}_i,
\end{equation}
where $\mathbf{S}_i = (S_i^x, S_i^y, S_i^z)$ represents pseudospin operators at site $i$.
For simplicity, only the interactions among the nearest-neighbor spins are considered.
In the first two terms, $K$ and $\Gamma$ are bond-directional exchange couplings of the Kitaev and $\Gamma$ terms, respectively.
For each bond, we can indicate an Ising axis $\gamma$ and label the bond as $\alpha\beta(\gamma)$,
with $\alpha$ and $\beta$ representing the other two remaining components.
Beyond the cubic $\{\textbf{e}_x, \textbf{e}_y, \textbf{e}_z\}$ axis, there is a relevant crystallographic $\{\textbf{a}, \textbf{b}, \textbf{c}\}$ frame
in which $\textbf{a} = (-\textbf{e}_x+\textbf{e}_y)/\sqrt2$, $\textbf{b} = (-\textbf{e}_x-\textbf{e}_y+2\textbf{e}_z)/\sqrt6$,
and $\textbf{c} = (\textbf{e}_x+\textbf{e}_y+\textbf{e}_z)/\sqrt3$.
In what follows, we will stick to the $\{\textbf{a}, \textbf{b}, \textbf{c}\}$ coordinate system [see the inset in Fig.~\ref{FIG-TmXPattern}(a)],
and the honeycomb lattice lies in the $\mathbf{a}$-$\textbf{b}$ plane.
The last term in Eq.~\eqref{Eq:HamKGH} represents the magnetic field and its direction is perpendicular to honeycomb plane, i.e., $\textbf{h} = h\textbf{c}$.
In this work, we parameterize $(K, \Gamma) = \mathcal{E}_0(\cos\phi, \sin\phi)$ and let $h$ varies freely.

To obtain configurations of classical ground states ($S \rightarrow \infinity$),
we perform the parallel-tempering Monte Carlo simulations in combination with heat-bath updates and over-relaxation methods \cite{HukushimaNemoto1996}.
After the Monte Carlo simulations, the classical ground state configurations can be obtained by iteratively aligning the spins with their local fields \cite{Janssen2016}.
The static structure factor is given by
$\mathcal{S}_{\mathbf{q}} = \frac{1}{N^2}\sum_{ij}\mathbf{S}_i\cdot \mathbf{S}_j e^{\imath \mathbf{q} \cdot (\mathbf{R}_i - \mathbf{R}_j)}$,
where $N$ is the number of sites and $\mathbf{R}_i$ is the location of spin at site $i$.

Then, we use the linear spin-wave theory to consider magnon excitations in the magnetically ordered state.
It is implemented by the the Holstein-Primakoff approximation.
When there are multiple spins in a magnetic unit cell, the spin at site $i$ can be expressed as \cite{Toth2015}
\begin{equation}\label{Eq:Quant}
\mathbf{S}_i = \sqrt{\frac{S}{2}}( \mathbf{\bar{u}}_i b^{\dagger}_i + \mathbf{{u}}_i b_i ) + \mathbf{{v}}_i(S - b^{\dagger}_i b_i),
\end{equation}
where $b^{\dagger}_i$ ($b_i$) is bosonic creation (annihilation) operator.
The auxiliary vector $\mathbf{{v}}_i$ is the classical spin direction
$\mathbf{{v}}_i = \mathbf{S}_i/S = (\sin\theta_i \cos\phi_i, \sin\theta_i \cos\phi_i, \cos\theta_i)$
while vector $\mathbf{{u}}_i$ can be calculated by
$\mathbf{{u}}_i = (\cos\theta_i \cos\phi_i - \imath \sin\phi_i, \cos\theta_i \sin\phi_i  + \imath \cos\phi_i, -\sin\theta_i)$.
After all spins within the magnetic unit cell are quantized, we can write the Hamiltonian in the reciprocal space as follows \cite{Janssen2019}
\begin{equation}\label{Eq:HamBDG}
\mathcal{H}=S(S+1) \mathcal{E}_0 E_g + \frac{\mathcal{E}_0 S}{2}\sum_{\mathbf{q}}\Psi_{\mathbf{q}}^{\dag}\mathcal{H}_{\mathbf{q}}\Psi_{\mathbf{q}}.
\end{equation}
Here, $S^2\mathcal{E}_0 E_g$ in the first part of Eq.~\eqref{Eq:HamBDG} is the classical ground-state energy,
while the second part stands for the quantum fluctuations due to magnons.
$\Psi_{\mathbf{q}}=(\bm{b}_{\mathbf{q}}, \bm{b}_{-\mathbf{q}}^{\dag})^T$,
$\bm{b}_{\mathbf{q}} = (b_{1\mathbf{q}}, \dots, b_{\mathcal{N}\mathbf{q}})$
and $\mathcal{N}$ is the number of spins in one magnetic unit cell.
Thus, the matrix $\mathcal{H}_{\mathbf{q}}$ can be divided into four blocks,
\begin{equation}\label{Eq:HamBDG2}
\mathcal{H}_{\mathbf{q}}=\left[\begin{matrix}\mathbf{A_{\mathbf{q}}} & \mathbf{B_{\mathbf{q}}}\\
\mathbf{B_{-\mathbf{q}}^{*}} & \mathbf{A_{-\mathbf{q}}^{T}}
\end{matrix}\right],
\end{equation}
where $\mathbf{A_{\mathbf{q}}}$ and $\mathbf{B_{\mathbf{q}}}$ are both $\mathcal{N}$-dimensional matrices.
Of note is that the contribution of the magnetic field has been included in $E_g$ and $\mathcal{H}_{\mathbf{q}}$.
For example, it is added to each diagonal element in the form of $-h \cos \theta_i/\left(\mathcal{E}_0 S\right)$ in the latter.
The $\mathcal{H}_\mathbf{q}$ is diagonalized by the Bogoliubov transformation,
\begin{equation}\label{Eq:BTrans}
\Psi_{\mathbf{q}}^{\dag}\mathcal{H}_{\mathbf{q}}\Psi_{\mathbf{q}} = \Psi_{\mathbf{q}}^{\dag}(\mathcal{T}^{-1}_{\mathbf{q}})^{\dagger}[\mathcal{T}^{\dagger}_{\mathbf{q}} \mathcal{H}_{\mathbf{q}} \mathcal{T}_{\mathbf{q}}]\mathcal{T}^{-1}_{\mathbf{q}}\Psi_{\mathbf{q}} = \Phi_{\mathbf{q}}^{\dag}E_\mathbf{q}\Phi_{\mathbf{q}},
\end{equation}
where $\mathcal{T}_\mathbf{q}$ is the transform matrix and
\begin{equation}\label{Eq:dispersions}
E_\mathbf{q}=\mathrm{diag}\big(E_{1,\mathbf{q}},\ldots,E_{\mathcal{N},\mathbf{q}},E_{1,-\mathbf{q}},\ldots,E_{\mathcal{N},-\mathbf{q}}\big)
\end{equation}
contains the magnon dispersions. Since $\Phi_{\mathbf{q}} $ can also be divided into two parts
$\Phi_{\mathbf{q}}=( \bm{\beta}_{\mathbf{q}} , \bm{\beta}_{-\mathbf{q}}^{\dag} )^T$
where $\bm{\beta}_{\mathbf{q}} = (\beta_{1\mathbf{q}},\dots,\beta_{\mathcal{N}\mathbf{q}})$,
the Bogoliubov transformation has a more detailed form \cite{Daniel2019},
\begin{equation}\label{Eq:BTrans2}
\begin{pmatrix} \bm{b}_{\mathbf{q}}   \\ \bm{b}_{-\mathbf{q}}^{\dag}   \end{pmatrix}  = \mathcal{T}_{\mathbf{q}} \begin{pmatrix} \bm{\beta}_{\mathbf{q}}   \\ \bm{\beta}_{-\mathbf{q}}^{\dag}   \end{pmatrix}
= \begin{pmatrix} U_{\mathbf{q}} & V_{-\mathbf{q}}^* \\ V_{\mathbf{q}} & U_{-\mathbf{q}}^*    \end{pmatrix} \begin{pmatrix} \bm{\beta}_{\mathbf{q}}   \\ \bm{\beta}_{-\mathbf{q}}^{\dag}   \end{pmatrix}.
\end{equation}

We can obtain the Berry curvature of the $n$-th energy band with the help of $\mathcal{T}_{\mathbf{q}}$ matrix
\begin{equation}\label{Eq:Berry}
\mathbf{\Omega}_{n\mathbf{q}} = -2 {\rm Im} \sum_{\substack{m=1\\m\neq n}}^{2\mathcal{N}} \frac{(\boldsymbol{\Sigma} \mathcal{T}^\dagger_k \partial_x \mathcal{H}_{\mathbf{q}} \mathcal{T}_k)_{nm}
	(\boldsymbol{\Sigma} \mathcal{T}^\dagger_k \partial_y \mathcal{H}_{\mathbf{q}} \mathcal{T}_k)_{mn}}{[(\boldsymbol{\Sigma}E_\mathbf{q})_{mm} - (\boldsymbol{\Sigma}E_\mathbf{q})_{nn}]^2},
\end{equation}
where $\boldsymbol{\Sigma} =\mathrm{diag}(\mathds{1}_{\mathcal{N}\times \mathcal{N}},-\mathds{1}_{\mathcal{N}\times \mathcal{N}}) $.
The Chern number of magnon band $n$ is the sum of Berry curvature in first Brillouin zone,
\begin{align}
\mathcal{C}_n = \frac{1}{2\pi} \int_{\mathbf{q}\in\text{FBZ}} \mathbf{\Omega}_{n\mathbf{q}} d^2\mathbf{q}.
\end{align}

\section{Results}
In our previous work \cite{Ken2023}, it is revealed that the TmX can be realized in the dominant $\Gamma$ region
and stabilized by the negative single-ion anisotropy or an out-of-plane magnetic field in the Kitaev-$\Gamma$ model.
The typical spin configuration of the TmX is shown in Fig.~\ref{FIG-TmXPattern}(a),
and the gray area containing eighteen spins marks the magnetic unit cell ($\mathcal{N} = 18$).
It displays an intricate pattern in which the core spins point along the $\textbf{c}$-axis while the surrounding spins lie almost in the honeycomb plane.
Figure~\ref{FIG-TmXPattern}(c) presents the corresponding static structure factor of the TmX, and a distinct ordering wave vector located at $2\mathbf{M}/{3}$ point is observed.
Of note is that such an interesting order belongs to the degenerate manifold of the classical honeycomb $\Gamma$ model,
and its spin-wave energy is surprisingly equal to that of the four-sublattice zigzag order \cite{LuoNPJ2021}.
The magnon band structure along the high-symmetry points depicted in Fig.~\ref{FIG-TmXPattern}(b) is shown in Fig.~\ref{FIG-TmXPattern}(d),
and the color in each band stands for the normalized Berry curvature.
It is observed that the lowest magnon band acquires a sizeable excitation gap at the $\bm{\Gamma}$ point,
and the lowest three magnon bands are well separated from the others with a global band gap.
Further, as will be shown later, at least some of total Berry curvature in the magnon bands does not cancel out, indicating that topologically nontrivial Chern number exists.
In what follows, we aim to unveil the topological phase transitions within the TmX.
Topological signatures such as the thermal Hall conductivity and chiral edge modes are also studied.

\begin{figure}[!ht]
  \centering
  \includegraphics[width=0.99\columnwidth, clip]{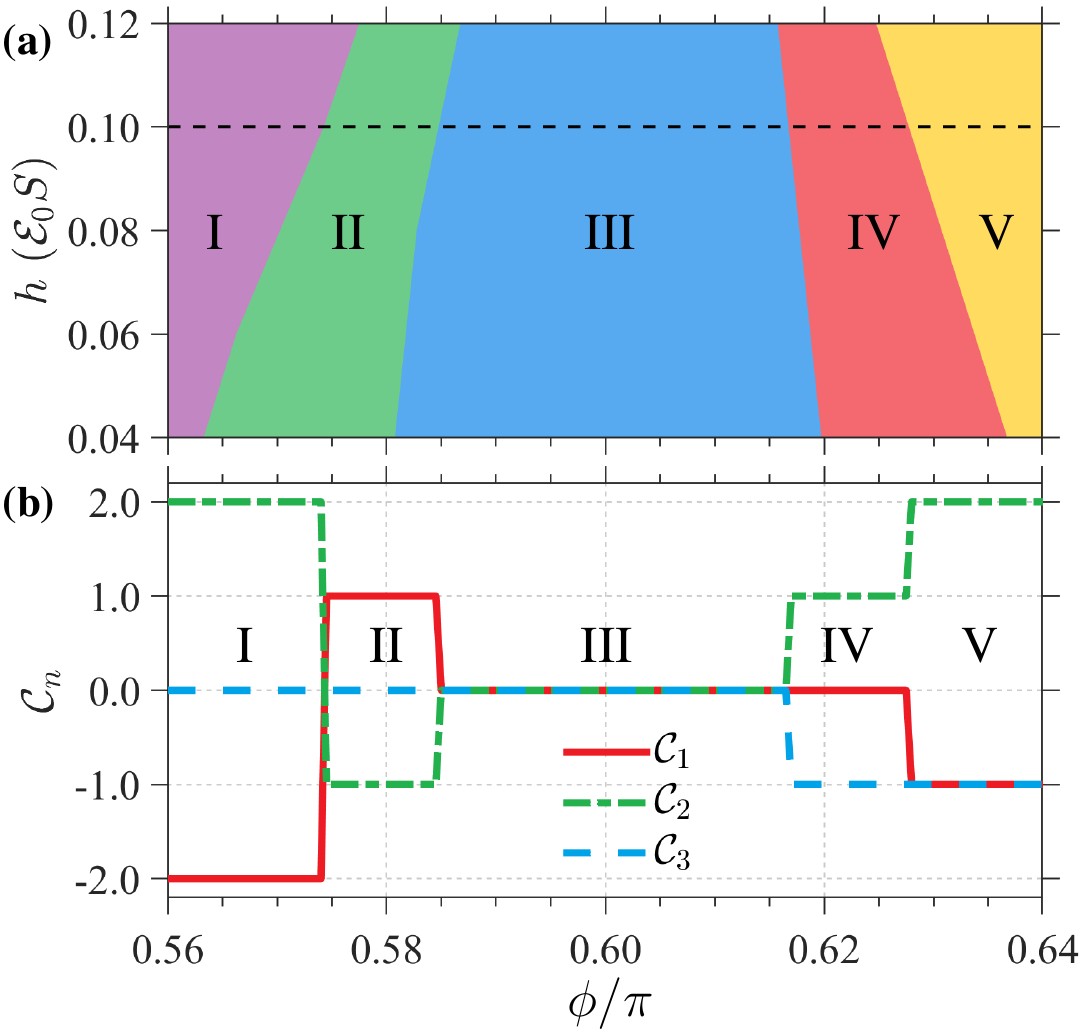}\\
	\caption{(a) Topological phase diagram in the $\phi$-$h$ plane.
    Five topological phases are distinguished through the Chern numbers
    and we distinguish them with Roman numerals I-V separately.
	(b) Behavior of the Chern numbers of the three lowest bands when $h/\left(\mathcal{E}_0 S\right) = 0.1$.
    }\label{FIG-PhaseDiag}
\end{figure}

\begin{figure*}[htb]
  \centering
\includegraphics[width=0.95\textwidth]{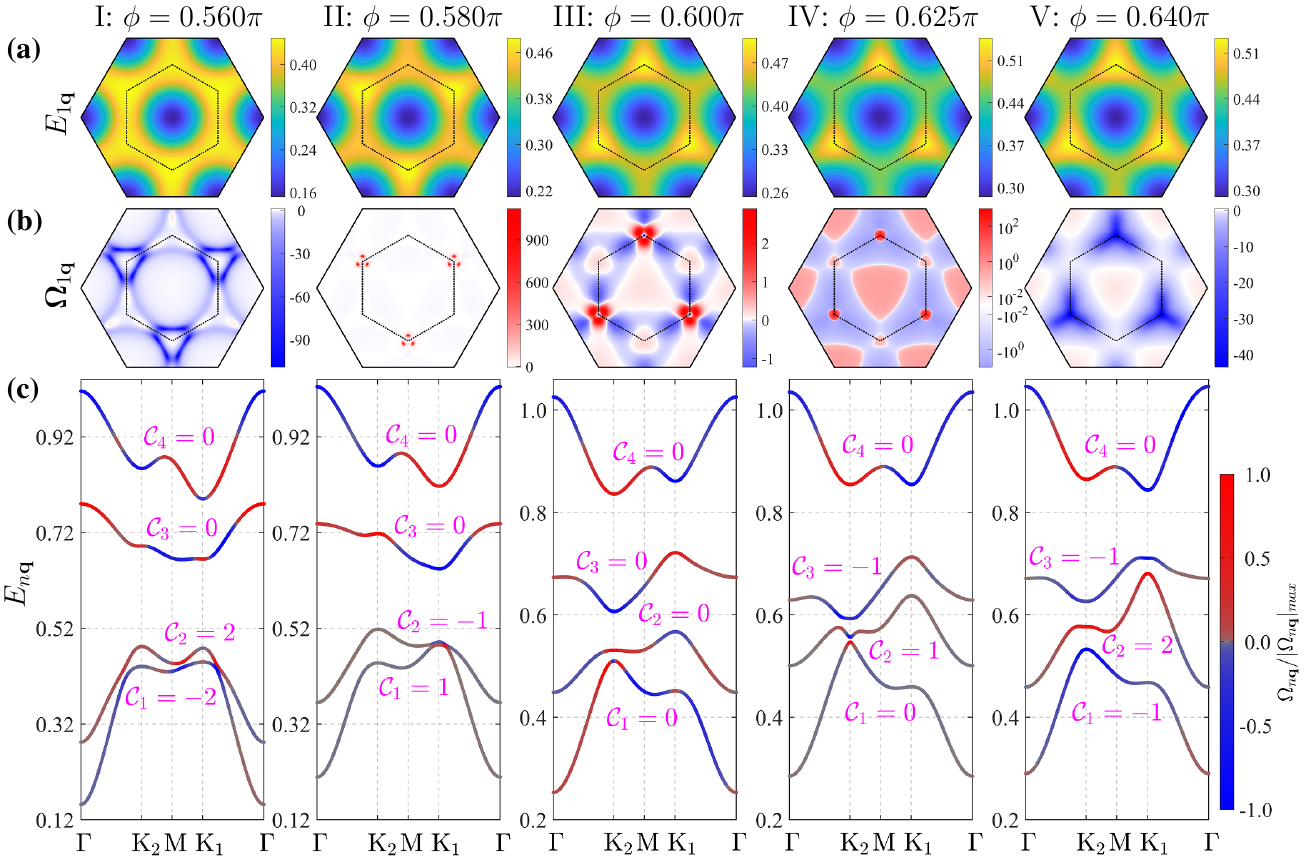}\\
    \caption{(a) and (b) show the typical spin-wave dispersions $E_{n{\bf{q}}}$ and the Berry curvatures $\Omega_{n\mathbf{q}}$ of each phase when $h/\left(\mathcal{E}_0 S\right) = 0.1$.
    Here, the lowest magnon bands with $n = 1$ are considered.
    (c) The band structure of the four lowest bands. The colors stand for the normalized Berry curvature and the Chern number is indicated for each band.
    }\label{FIG-Dispersion}	
\end{figure*}

\subsection{Topological phase diagram}
Despite great efforts, the notorious difficulty in mapping out the ground-state phase diagram of the Kitaev-$\Gamma$ model remains unsolved even at the classical level (for a review, see Ref.~\cite{Rousochatzakis2023}).
However, armed with advanced Monte Carlo methods, there has been a consensus on the recognition of the $C_3$-like TmX stemming from the dominant $\Gamma$ region.
This phase is relatively stable against ferromagnetic Kitaev interaction and extends to a large regime in the presence of bond/single-ion anisotropy \cite{Rayyan2022,Ken2023}.
Meanwhile, we identify that the TmX can survive on the existence of a small out-of-plane magnetic field,
and the fact that the magnetic field can open up the band gap is beneficial for the occurrence of topological magnons.
Due to the competition between the Kitaev and $\Gamma$ interactions, as well as the enhancement of magnetic field,
it leaves the possibility of topological phase transitions within the wide regime of TmX.

The Chern number associated with the Berry curvature is the most prominent quantity to capture topological phase transitions,
which manifests itself by the change in value.
Figure~\ref{FIG-PhaseDiag}(a) shows the topological phase diagram of the TmX in the range of $\phi/\pi \in [0.56, 0.64]$ and $h/\left(\mathcal{E}_0 S\right) \in [0.04, 0.12]$.
There are at least five distinct topological phases which have different sets of Chern numbers for the full magnon bands.
Specifically, sets of the Chern numbers ($\mathcal{C}_1$, $\mathcal{C}_2$, $\mathcal{C}_3$) of the lowest three magnon bands are
($-2, 2, 0$), ($1, -1, 0$), ($0, 0, 0$), ($0, 1, -1$), and ($-1, 2, -1$) for the phases ranging from I to V.
We note that the phase III is indeed topological as its Chern number of the fifth magnon band is nonzero.
To further affirm the existence of topological phase transitions,
we present the behaviors of Chern numbers in the lowest three magnon bands as a function of $\phi$ at $h/\left(\mathcal{E}_0 S\right)$ = 0.1, see Fig.~\ref{FIG-PhaseDiag}(b).
These curves clearly demonstrate successive topological phase transitions via the change of Chern numbers.
In addition to topological phase transitions, we emphasize that there is no magnetic phase transition within the regime of TmX.
This can be seen by the facts that the first and second derivatives of the ground-state energy do not show any singularity,
and intensities of the out-of-plane moment and static structure factor at $2\mathbf{M}/{3}$ point are smoothly changed as $\phi$ varies
(for illustration, see Appendix \ref{appendixA}).
Therefore, similar to the spin-flop phase identified in the extended Kitaev model \cite{Mook2022},
our work provides another example where the topological phase transitions can occur even though the magnetic phase transition is absent.

The nonreciprocal magnons come from the spatial inversion symmetry breaking \cite{Takuya2020,Satoru2022}
and is known to be stabilized by the dipole-dipole anisotropy, the Dzyaloshinskii-Moriya interaction, the symmetric anisotropic exchange, etc \cite{Satoru2022}.
The nonreciprocal magnon dispersions are stated as $E_{n, {\bf{q}}} \neq E_{n, -{\bf{q}}}$,
whereby magnons at momentum ${\bf{q}}$ have different energy from those at $-{\bf{q}}$.
They can be detected experimentally in LiFe$_5$O$_8$ and $\alpha$-Cu$_2$V$_2$O$_7$,
and the relevant physical phenomena such as nonreciprocal optical response and nonreciprocal spin Seebeck effect are also studied theoretically \cite{Takuya2020}.
In this regard, it is naturally to ask if the nonreciprocal magnons occur in the TmX.
The configuration in the TmX has double degeneracy that the vertical spins are at the different sublattices \cite{Ken2023}.
The nonreciprocal magnons are expected to appear since any of the degenerate configurations breaks the sublattice symmetry.

Figure~\ref{FIG-Dispersion} shows the typical spin-wave dispersions $E_{n{\bf{q}}}$ and the Berry curvatures $\Omega_{n\mathbf{q}}$ of the lowest magnon band ($n = 1$).
As can be seen from Fig.~\ref{FIG-Dispersion}(a), the nonreciprocal magnons are clearly reflected in the relation $E_{1, {\bf{q}}} \neq E_{1, -{\bf{q}}}$.
This asymmetry is demonstrated by the energy difference between the neighboring points at the corners of the first Brillouin zone,
i.e., $\delta E_n = \vert E_{n{\bf K}_1} - E_{n{\bf K}_2} \vert$.
It is found that $\delta E_n$ is finite throughout the regime of TmX, advocating the existence of nonreciprocal magnons.
Interestingly, the kinks in the curves of $\delta E_n$ ($n$ = 1, 2, 3) are coincident with the topological phase transitions, see Appendix \ref{appendixB}.
Further, as shown in Fig.~\ref{FIG-Dispersion}(b), the Berry curvatures $\Omega_{1, \mathbf{q}}$ are mostly concentrated around the points pertaining band gaps.
However, distributions of the Berry curvatures throughout the first Brillouin zone are rather distinct among different phases.
To begin with, values of the Berry curvature in phases I, II, or V are overwhelmingly negative or positive, leading to a finite Chern number ultimately.
The Chern numbers of these three regimes are $-2$, $1$, and $-1$, respectively.
Nevertheless, the Chern numbers in phase III and IV are zero but their reasons are different.
In phase III, both area and intensity of the Berry curvature in the negative and positive regimes are close, and the Chern number is thus zero.
By contrast, in phase IV, although intensities of the Berry curvature at corners of the first Brillouin zone are extremely large,
area of the regime of positive Brillouin zone is so small that it cancels with that of the negative counterpart.
Finally, the first four magnon bands, together with the individually normalized Berry curvature, are shown in Fig.~\ref{FIG-Dispersion}(c).
It is found that the fourth magnon bands of these topological phases always acquire a zero Chern number and are well separated from the lowest three with global band gaps.
Recalling that magnons follow the Bose-Einstein distribution, topological quantities of magnons should highly rely on the lowest energy bands at low temperature.
The above fact thus highlights the importance of the role played by the lowest three bands.
Also, the intricate relation between the Berry curvature and magnon energy accounts for the elusive behaviors of various topological quantities.

\begin{figure}[htpb]
  \centering
  \includegraphics[width=0.9\columnwidth]{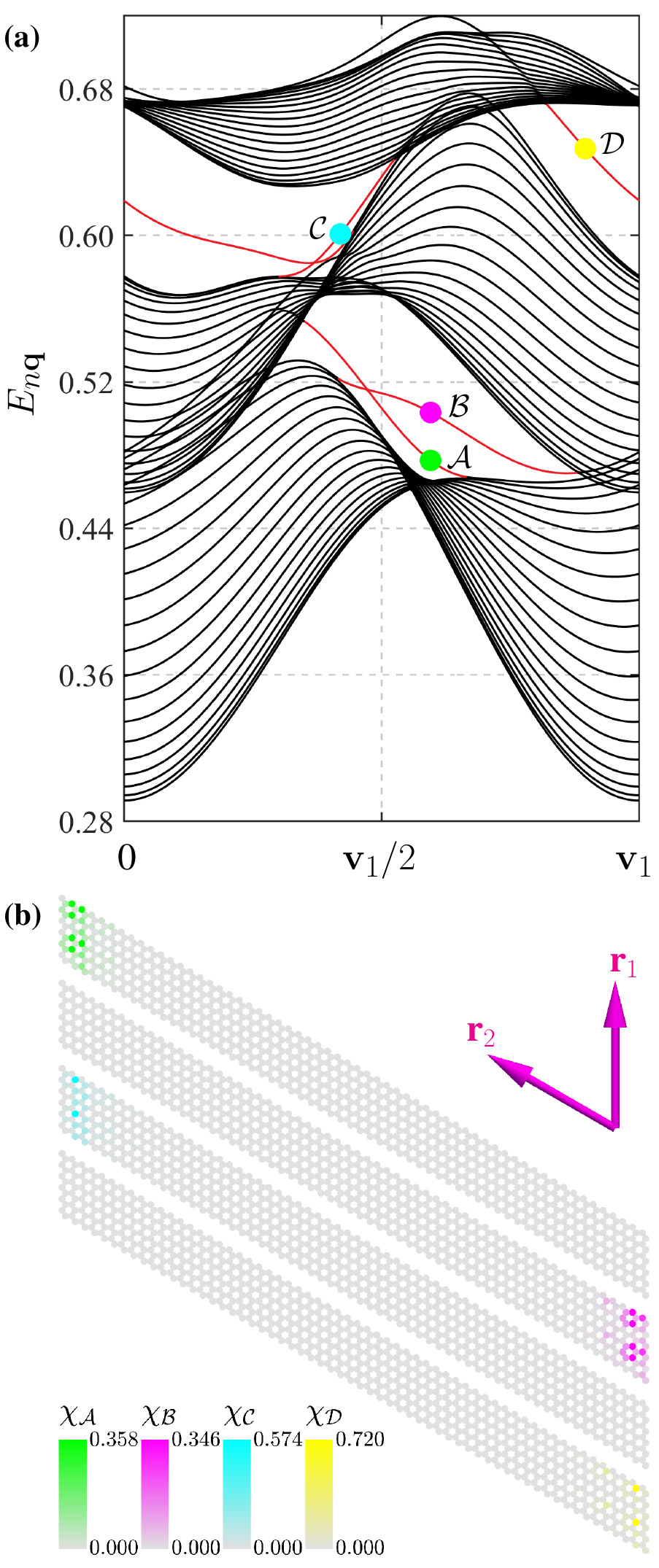}
  \caption{(a) The magnon spin-wave dispersions on a nanoribbon geometry at $\phi = 0.64\pi$ and $h/\left(\mathcal{E}_0 S\right) = 0.1$.
  The energy bands that contain chiral edge modes are depicted in red.
  The periodic boundary condition is used along $\textbf{r}_1$ direction while open boundary condition is used along $\textbf{r}_2$ direction.
  The length in $\textbf{r}_2$ direction is 20 times the primitive vector while in $\textbf{r}_1$ direction it equals to the primitive vector.
  The $\textbf{v}_1$ in the horizontal axis is the inverse primitive vector of $\textbf{r}_1$.
  (b) The intensities reveal the real space magnonic contribution $\chi_i {\left(\mathbf{q}\right)}$ at four representative points (labeled as $\mathcal{A}$-$\mathcal{D}$ in (a)) with chiral edge modes. The $\textbf{r}_1$ direction is enlarged twice for better visual effect.
  }\label{FIG-EdgeModes}
\end{figure}

\subsection{Chiral edge modes}
The nontrivial band topology can be confirmed by calculating the chiral edge states in a nanoribbon geometry \cite{Mook2014}.
When the open boundary condition is adopted, there will be chiral edge states connecting the upper and lower energy bands.
According to the bulk-edge correspondence, the number of pairs of edge states in the $n$th band gap
is consistent with the winding number $\mathcal{W}_n = \sum_{m \leq n}\mathcal{C}_n$ \cite{Hatsugai1993}.
We here consider the phase V since it is the only case where both $\mathcal{W}_1$ ( = $-1$) and $\mathcal{W}_2$ ( = $+1$) are nonzero.
However, in contrast to the well recognized cases where global band gaps exist,
the first two global band gaps are absent [as shown in Fig.~\ref{FIG-TmXPattern}(d)], challenging the capture of well-defined edge states.
Figure~\ref{FIG-EdgeModes}(a) shows the magnon band structure on a nanoribbon geometry at $\phi = 0.64\pi$ and $h/\left(\mathcal{E}_0 S\right) = 0.1$.
Strictly speaking, at the edge of the open boundary, the configuration is no longer a perfect TmX shape, but the influence of the boundary will decrease as the system size increases. Therefore, we ignore the influence of boundary conditions on the configuration, and construct the configuration of nanoribbon geometry by translating a single TmX magnetic unit cell.
We manage to identify a pair of additional bands (shown as red lines) connecting the upper and lower bulk bands.
These four bands are mixed with the bulk ones and are only distinguishable at certain momentum intervals,
leaving the possibility to study the chiral edge modes thereof.

We proceed to reveal the magnonic contribution of the wave functions of these chiral edge modes in real space.
Following the definition proposed in Ref.~\cite{Daniel2019}, the magnonic contribution at site $i$ is given by
\begin{align}\label{Eq:Chi}
\chi_{i}(\mathbf{q}) = |\langle GS|b_{i\mathbf{q}} \beta_{n\mathbf{q}}^{\dagger}|GS\rangle|^2 = |U_{\mathbf{q}}^{i,n}|^2,
\end{align}
where $\beta_{n\mathbf{q}}^{\dagger}|GS\rangle$ is the single magnon state
and $|GS\rangle$ is the ground state that satisfies $\beta_{n\mathbf{q}}|GS\rangle = 0$.
The matrix element $U_{\mathbf{q}}^{i,n}$ is the part of the transformation matrix $\mathcal{T}_{\mathbf{{q}}}$, see Eq.~\eqref{Eq:BTrans2}.
Figure~\ref{FIG-EdgeModes}(b) shows the magnonic contribution of four representative points labeled as $\mathcal{A}$-$\mathcal{D}$ in Fig.~\ref{FIG-EdgeModes}(a).
It is thus clear that magnons are localized at different edges of the nanoribbons, indicating these additional bands are indeed chiral edge modes.
In addition, we also calculate the magnonic contribution of other bands that are equipped with the same band energy of the individual points at $\mathcal{A}$-$\mathcal{D}$.
It is observed that $\chi_{i}(\mathbf{q})$ is almost uniformly distributed in the nanoribbon geometry and their intensities are rather small,
advocating the nontrivial properties of the chiral edge modes shown in Fig.~\ref{FIG-EdgeModes}(b).

\subsection{Thermal Hall effect}
Upon applying a longitudinal temperature gradient, the nonzero Berry curvature can carry a transverse heat current, leading to the magnon thermal Hall effect.
It is manifested by a nonzero thermal Hall conductivity ($\kappa_{ab}$) defined as \cite{Matsumoto2014},
\begin{align}\label{Eq:Kxy}
    \kappa_{ab} = -\frac{k_B^2 T}{ 4\pi^2 \hbar}
	\sum_{n=1}^{\mathcal{N}}  \int_{\mathbf{q} \in \text{FBZ}}
	c_2\left[\rho\left(E_{n\mathbf{q}}\right)\right] \mathbf{\Omega}_{n\mathbf{q}} d^2\mathbf{q},
\end{align}
where $\rho(E_{n\mathbf{q}}) = 1/\left(e^{\mathcal{E}_0 SE_{n\mathbf{q}}/k_B T}-1\right)$ is the Bose-Einstein distribution.
The weighting function $c_2(x)=(1+x)\ln^2[(1+x)/x] - \ln^2(x) - 2\text{Li}_2(-x)$, with $\text{Li}_2(x)$ being the polylogarithm function.

\begin{figure}[!ht]
  \centering
  \includegraphics[width=1\columnwidth]{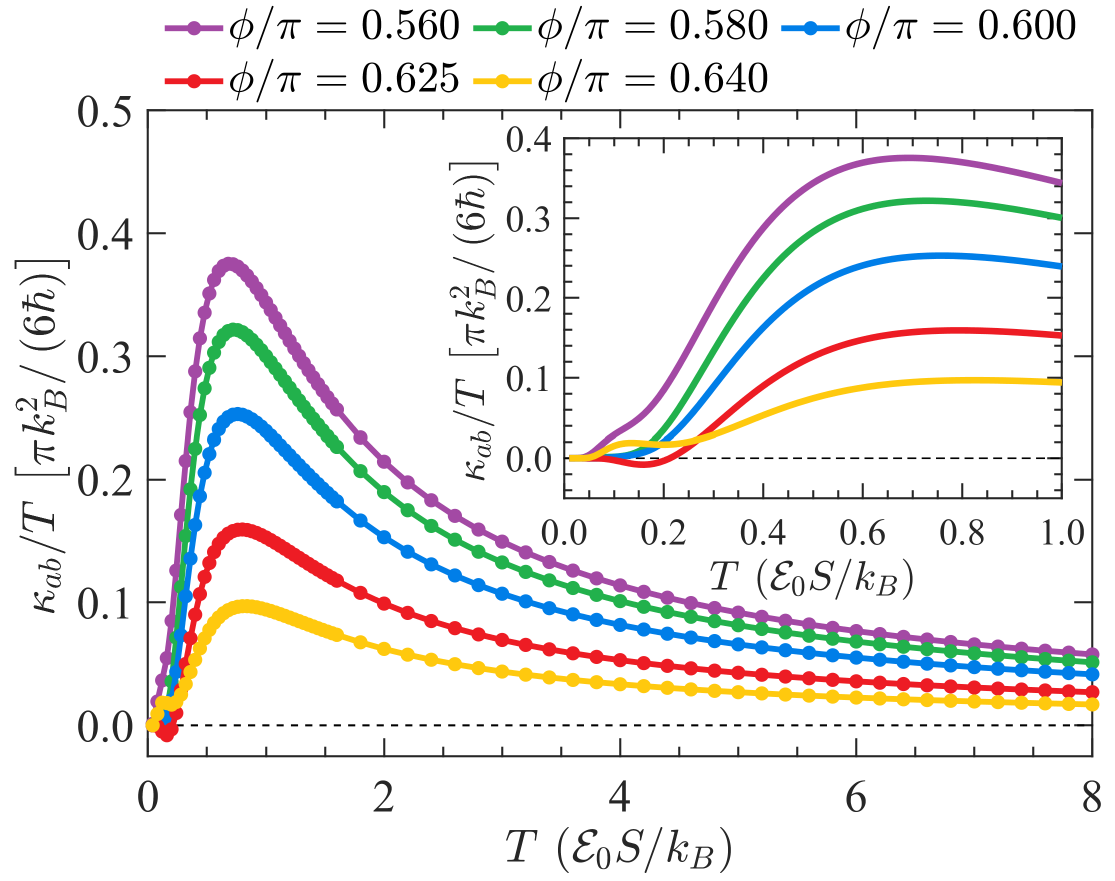}\\
  \caption{$\kappa_{ab}/T$ (in units of $\pi k_B^2/\left(6\hbar\right)$)  as a function of $T$ for different $\phi$ that belong to phase I-V separately.
  The inset shows the magnified results of the low-temperature.
  In region with relatively high temperature up to $k_BT/\left(\mathcal{E}_0 S\right) \lesssim 8$, 
  we consistently ignore the influence of thermal fluctuations on magnetic configurations and the magnon-magnon interactions.}
  \label{FIG-ThermalHallKxydT}	
\end{figure}

Figure~\ref{FIG-ThermalHallKxydT} shows behaviors of $\kappa_{ab}/T$ at five selected points in each distinct topological phase.
In the high-temperature limit,
$\kappa_{ab}$ saturates to the value of \cite{Mook2014b}
\begin{align}\label{Eq:KxyTInf}
\kappa_{ab}^{\lim} = frac{\mathcal{E}_0 S k_B}{ 4\pi^2 \hbar}
\sum_n  \int_{\mathbf{q} \in \text{FBZ} }E_{n\mathbf{q}}\mathbf{\Omega}_{n\mathbf{q}} d^2\mathbf{q},
\end{align}
indicating that $\kappa_{ab}/T$ obeys the law of $\propto T^{-1}$ at sufficiently large temperature.
As inferred from Eq.~\eqref{Eq:KxyTInf}, the saturation value depends on the distributions of dispersion relation and Berry curvature of each band.
It is observed from Fig.~\ref{FIG-ThermalHallKxydT} that the saturation value decreases with the increase of $\phi$.
Further, $\kappa_{ab}/T$ displays a pronounced peak at each curve.
The magnitudes (in unit of $\pi k_B^2/\left(6\hbar\right)$) are smaller than $1/2$, the half-quantized value in the case of majorana fermion.
By contrast, positions of these peaks are insensitive to $\phi$ and are close to $k_BT/\left(\mathcal{E}_0 S\right) \approx 0.7$.
It is interesting to note that the energy scale of this temperature falls in the global band gap
that separates the lowest three magnon band with others [see Fig.~\ref{FIG-TmXPattern}(d)].
This result demonstrates that the former plays a vital role in the low-temperature thermal Hall conductivity.
As seen from inset of Fig.~\ref{FIG-ThermalHallKxydT}, $\kappa_{ab}/T$ opens up exponentially when $T$ is relatively small.
As  $T$ further increases, there are kind of enhancement of $\kappa_{ab}/T$ in the curves of $\phi/\pi = $ 0.56 (phase I) and 0.64 (phase V).
This may result from the higher Chern number of 2 existed in the lowest three magnon bands.
Noteworthily, when $\phi/\pi = $ 0.625 (phase IV), $\kappa_{ab}/T$ undergoes an appreciable sign change from negative to positive at $k_BT/\left(\mathcal{E}_0 S\right) \simeq 0.21$.
Notice that the first  magnon band is trivial, the negative thermal Hall conductivity at low temperature
is thus attributed to the second magnon band which owns a Chern number of $+1$.
The interplay of the lowest three magnon bands is shown in Appendix \ref{appendixC}.

\begin{figure}[!ht]
  \centering
  \includegraphics[width=1\columnwidth]{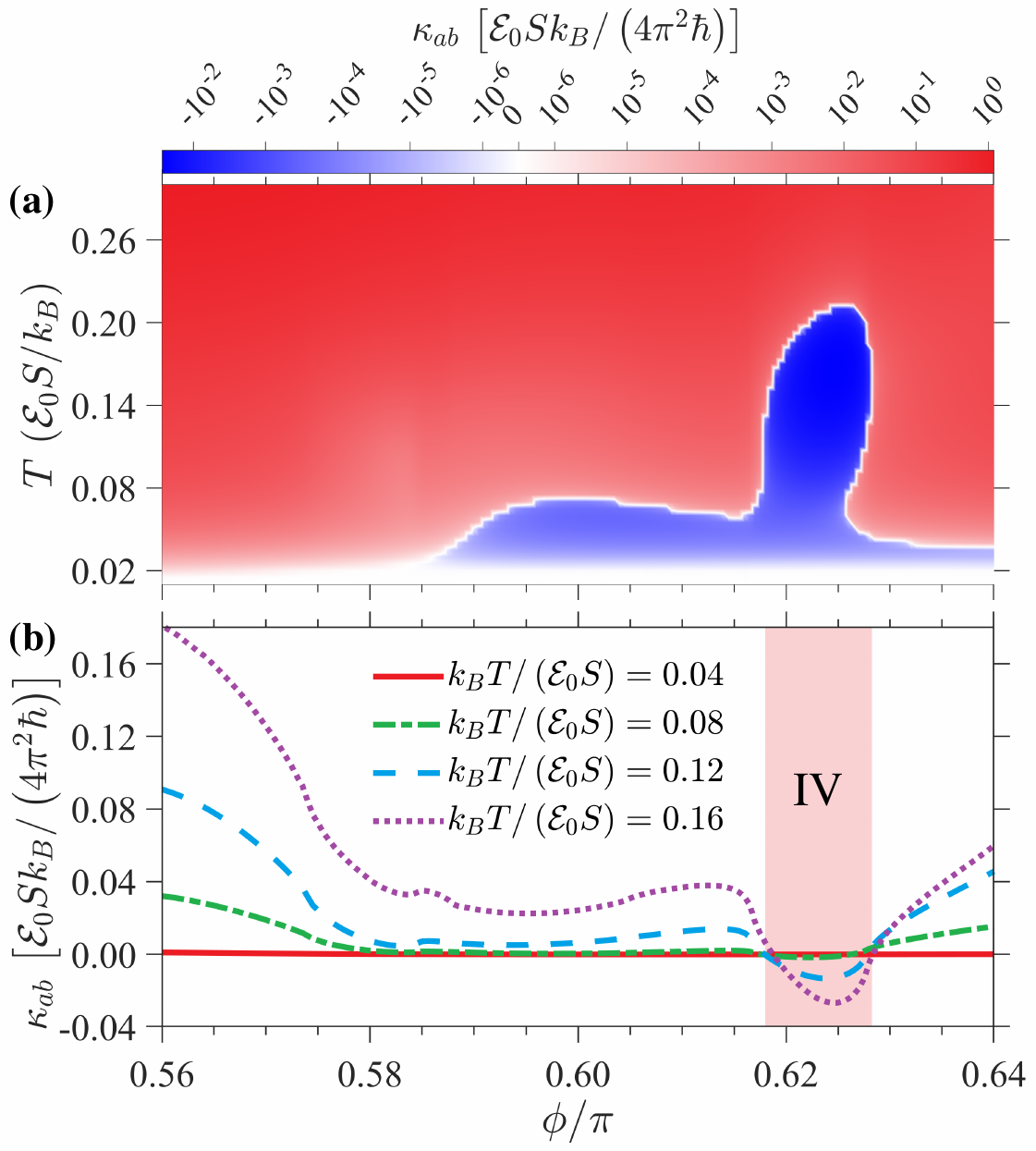}\\
	\caption{(a) The contour plot of $\kappa_{ab}$ in as a function of $T$ and $\phi$. (b) shows $\kappa_{ab}$ as a function of $\phi$ for four different temperatures.
	The shaded pink region indicates the range of phase IV.}	
	\label{FIG-ThermalHallSign}
\end{figure}

To better visualize the sign change in $\kappa_{ab}$ at low temperature,
we present the contour plot of $\kappa_{ab}$ in Fig.~\ref{FIG-ThermalHallSign}(a) as a function of $T$ in the regime of TmX. 
At low temperatures [e.g., $k_BT/\left(\mathcal{E}_0 S\right) = 0.02$], signs of $k_{xy}$ are basically positive in phases I and II,
while they are negative in phases III, IV, and V.
Since signs of $\kappa_{ab}$ in all the five phases are positive at large enough temperature,
the latter are expected to undergo sign changes as $T$ is further lifted.
Of note is that the sign change in phase IV is the most prominent and it remains negative when $k_BT/\left(\mathcal{E}_0 S\right) \lesssim 0.21$.
Figure~\ref{FIG-ThermalHallSign}(b) shows $\kappa_{ab}$ as a function of $\phi$ for four different temperatures.
At each temperature, $\kappa_{ab}$ changes nonmonotonously with $\phi$ and the unusual behaviors may have a plausible relation to the underlying topological phase transitions.
Specifically, it is observed that the sign of $\kappa_{ab}$ in phase IV is different from its neighboring phases.
For the selected temperatures $k_BT/\left(\mathcal{E}_0 S\right)$ = 0.08, 0.12, and 0.16,
locations of the sign changes are robust and coincide nicely with the phase boundaries of phase IV (see the shaded pink region).
Our result reveals that the sign change of thermal Hall conductivity is amenable to serve as a diagnosis of topological phase transitions.

\section{CONCLUSIONS}
In this paper, we have studied the topological phase transitions and nontrivial thermal Hall effect in a noncollinear spin texture termed triple-meron crystal (TmX).
It is discovered that the TmX occupies a large parameter region near the $\Gamma$ limit and is stabilized by the out-of-plane magnetic field
in the Kitaev-$\Gamma$ model through parallel-tempering Monte Carlo simulations.
Further, we obtain the magnon dispersions and Berry curvatures successfully with the help of the linear spin-wave theory,
from which the Chern number, chiral edge mode, and thermal Hall conductivity can be calculated.
Throughout the regime of TmX, we map out a topological phase diagram by the Chern number and identify five distinct topological phases therein.
Due to the existence of symmetric anisotropic exchanges, the topological magnons display nonreciprocal structures
and the behavior of nonreciprocity is helpful to reveal the underlying topological phase transitions.
The topological nature of magnons is also verified by the onset of chiral edge modes in a nanoribbon geometry.
We confirm that the pair of nontrivial edge states equals to that of the winding number at the corresponding band level.
Finally, we observe that the thermal Hall conductivity ($\kappa_{ab}$) enjoys a sign change at low temperature in some parameter region
and the peak of $\kappa_{ab}/T$ is modest and comparable to the half-quantized value due to majorana fermion.
Guided by the topological phase diagram, we can relate the sign change in $\kappa_{ab}$ to a certain species of topological phase transition.

The significance of our work lies in that it underscores topological magnons in a noncollinear spin texture stabilized by Kitaev interactions, thus it should illuminate future studies of bosonic topological band theory on Kitaev materials. In addition to the content presented in this article, there are still some issues worth further research. First of all, while we have predicated that the TmX can be realized in higher-spin Kitaev magnets, potential candidates are still lacking. We thus hope that our finding could stimulate the synthesis of proper materials so as to solidify the topological magnons.
Next, the magnon-magnon interactions may lead to the decay of quasiparticles \cite{Chernyshev2013,Mook2020prr,Mook2021prx,Koyama2023,Lu2021,Habel2024}
or make them more stable \cite{Verresen2019}.
Recent works have also pointed out that
the magnon-magnon interactions may have a promoting effect on the formation of band topology \cite{Mook2021prx,Gohlke2023}.
Hence, it is meaningful to further discuss the relevant fields based on our work.
Finally, since the phonons are omnipresent and play a crucial role in the low-energy thermal transport,
it is necessary to analyze the effect of spin-lattice coupling on the thermal Hall conductivity of certain materials \cite{Sun2023,Park2023,Ke2024}.

\begin{acknowledgments}
We would like to thank Satoru Hayami for helpful discussions.
This work is supported by the National Key R\&D Program of China (Grants No. 2022YFA1402704),
	by the National Natural Science Foundation of China (Grants No. 12274187, No. 12304176, No. 12247183, No. 12247101, No. 11834005),
and by the Natural Science Foundation of Jiangsu Province (Grant No. BK20220876).
The computations are partially supported by High Performance Computing Platform of Nanjing University of Aeronautics and Astronautics (NUAA).
\end{acknowledgments}



\appendix

\begin{figure}[!ht]
	\centering
	\includegraphics[width=0.98\columnwidth]{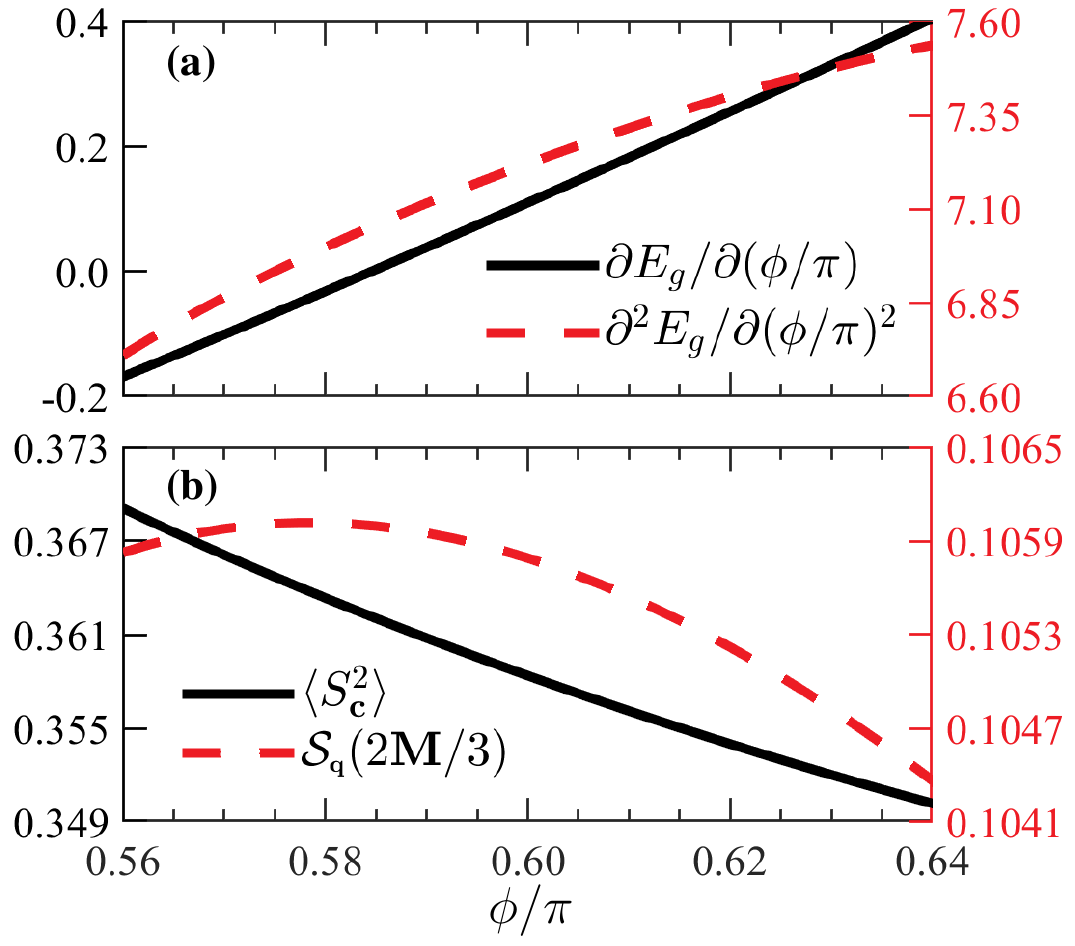}\\
	\caption{(a) The first-order $\partial E_g/ \partial (\phi/\pi) $ and second-order $\partial^2 E_g/\partial(\phi/\pi)^2$ derivatives of the ground-state energy per site.
		(b) The square of the out-of-plane component of spin  $\langle S_{\mathbf{c}}^{2} \rangle $
		and static structure factor $\mathcal{S}_{\textbf{q}}$ at $\textbf{q} = 2\mathbf{M}/{3}$ point.
	}\label{FIG-appendixA}
\end{figure}

\section{Absence of magnetic phase transition}\label{appendixA}
We note that there is no magnetic phase transition within the wide regime of TmX under a small magnetic field.
As a comparison, a series of topological phases are recognized in the TmX, see Fig.~\ref{FIG-PhaseDiag}(a).
In this appendix, we focus on the line $h/\left(\mathcal{E}_0 S\right) = 0.1$ as an example to confirm the absence of magnetic phase transition.
Figure~\ref{FIG-appendixA}(a) shows the first-order $\partial E_g/ \partial (\phi/\pi) $ and second-order $\partial^2 E_g/\partial(\phi/\pi)^2$ derivatives of the ground-state energy as a function of $\phi$.
These curves are smooth enough, ruling out a possibility of displaying kink, jump, or divergence.
Further, square of the out-of-plane component of spin $\langle S_{\mathbf{c}}^{2}\rangle$ and static structure factor $\mathcal{S}_{\textbf{q}}$ at $\textbf{q} = 2\mathbf{M}/{3}$ point are shown in Fig.~\ref{FIG-appendixA}(b).
They are also smoothly varied as $\phi$, indicating that the magnetic phase transition is unlikely to occur.
Taken together, it is safely to conclude that there is no magnetic phase transition in the regime of TmX.

\section{Evidence of nonreciprocal magnons}\label{appendixB}
\begin{figure}[!ht]
	\centering
	\includegraphics[width=1\columnwidth]{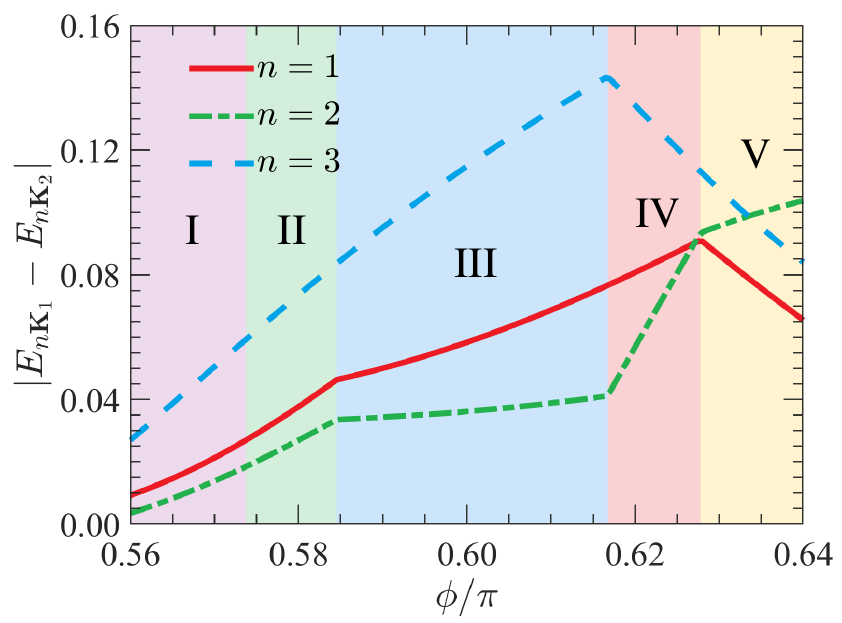}\\
	\caption{The difference in magnon energy at $\textbf{K}_1$ and $\textbf{K}_2$ points of band $n$ (= 1, 2, 3) as a function of $\phi$.}	
	\label{FIG-appendixB}
\end{figure}

The high symmetry points ${\bf K}_1$ and ${\bf K}_2$ have opposite positions $\mathbf{q}$ in reciprocal space.
According to the spin-wave dispersions shown in Fig.~\ref{FIG-Dispersion}(a),
the quantity $\delta E_n = \vert E_{n{\bf K}_1} -  E_{n{\bf K}_2}\vert$ serves as an indicator of non-reciprocity.
Figure~\ref{FIG-appendixB} shows $\delta E_n$ ($n$ = 1, 2, 3) as a function of $\phi$ when $h/\left(\mathcal{E}_0 S\right) = 0.1$.
Apparently, all the $\delta E_n$'s are finite and their values becomes larger and larger averagely as $n$ increases,
confirming the existence of nonreciprocal magnons.
In addition, it is interesting to note that $\delta E_n$'s have an implicit relation to the underlying topological phase transitions.
As can be seen from Fig.~\ref{FIG-Dispersion}(c), due to the nonreciprocity of magnons, the difference in band gaps between ${\bf K}_1$ and ${\bf K}_2$ points is significant.
When a phase transition occurs, the energy band only closes at one point among them,
and the Berry curvature corresponding to this point contributes the most to the Chern number.
This further leads to the connection between topological phase transitions and the quantity $\delta E_n = \vert E_{n{\bf K}_1} -  E_{n{\bf K}_2}\vert$.
For example,
$\delta E_1$ has two kinks at II-III and IV-V transitions,
$\delta E_2$ has three kinks at II-III, III-IV, and IV-V transitions,
while $\delta E_3$ has one kink at III-IV transition.
These kinks are precisely located at the topological phase transition points.
In addition, $\delta E_n$  do not show a kink at I-II transition.
This is because the closure point of the energy band is no longer the ${\bf K}_1$ or ${\bf K}_2$ point during such topological phase transition.

\section{Dissecting the thermal Hall conductivity in phase IV}\label{appendixC}
\begin{figure}[!ht]
	\centering
	\includegraphics[width=1\columnwidth]{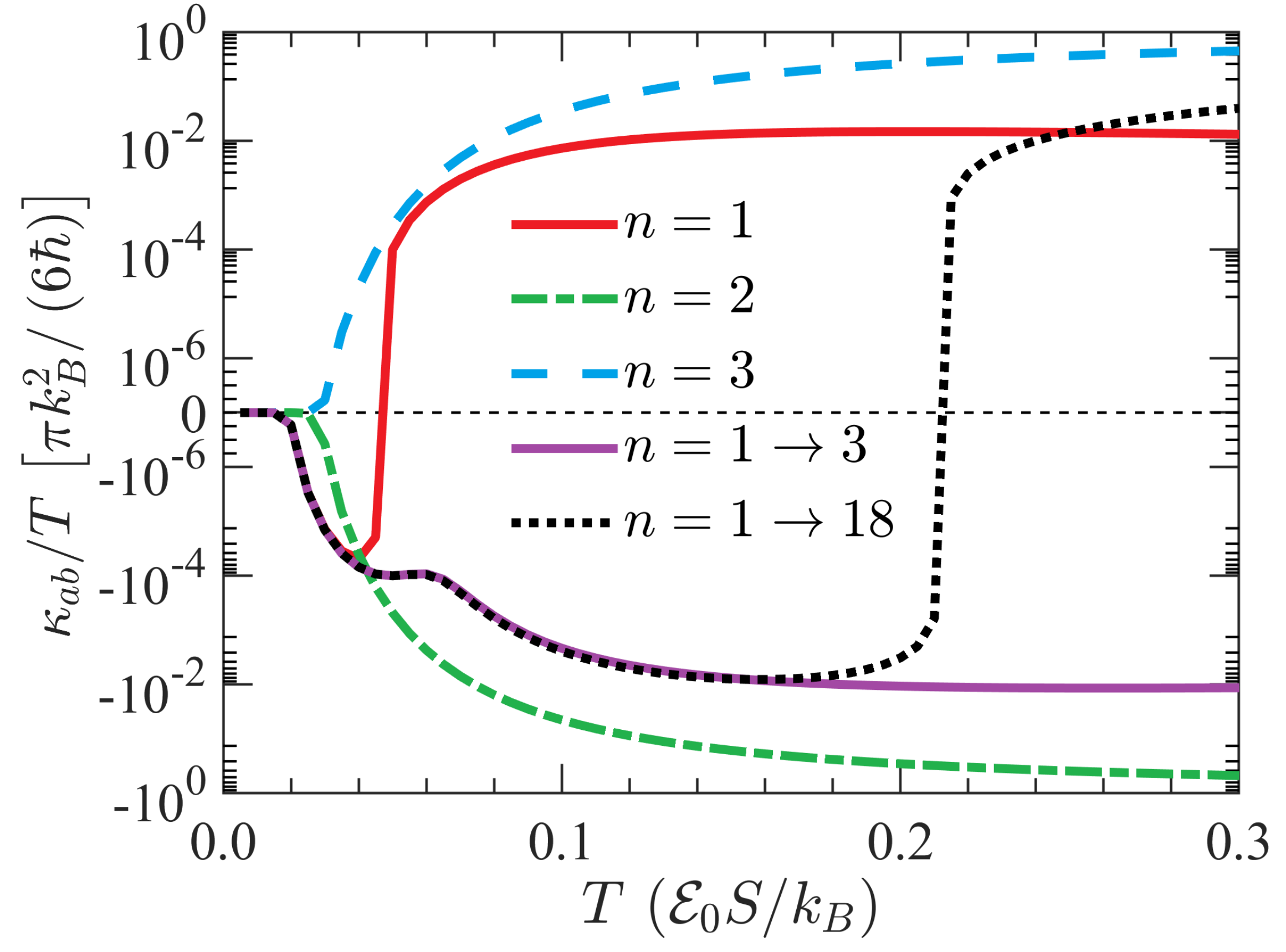}\\
	\caption{$\kappa_{ab}/T$ (in units of \textcolor{red}{$\pi k_B^2/\left(6\hbar\right)$})  as a function of \textcolor{red}{$T$ }at a representative point ($\phi = 0.625\pi$,\textcolor{red}{ $h/\left(\mathcal{E}_0 S\right) = 0.1$)} in  phase IV.
		The curves $n=$ 1, 2, 3 are the results coming from the individual band.
		Results of total three lowest bands $n=1\to3$ and all eighteen bands $n=1\to18$ are also provided.
	}\label{FIG-appendixC}
\end{figure}

As seen from Fig.~\ref{FIG-ThermalHallSign}(a),
the thermal Hall conductivity of phase IV has a sizable negative value when the temperature is small,
and it becomes positive as the temperature increases.
Since the lowest three magnon bands have a significant contribution to thermal Hall conductivity at low temperatures,
here we calculate their thermal Hall conductivity separately.
We recall that the first three Chern numbers in this phase are $(0, 1, -1)$.
As the Berry curvature cancels out in the first band,
it means that this band plays an insignificant role when compared with the remaining two.
As shown in Fig.~\ref{FIG-appendixC}, the contribution of first band is indeed tiny except at the low enough temperature
and is two orders smaller than that of the second and third bands when $k_BT/\left(\mathcal{E}_0 S\right) = 0.3$.
According to the definition in Eq.~\eqref{Eq:Kxy}, the sign of thermal Hall conductivity is generally opposing to its Chern number.
Thus, the signs of thermal Hall conductivity in the second and third bands are negative and positive, respectively.
Notably, it is \textcolor{red}{seen} that band 2 completely offsets the contributions of bands 1 and 3,
resulting in a negative thermal Hall conductivity in the low-temperature region (see the purple dotted line).
As a comparison, we also present the total contribution of all eighteen bands,
and the two curves are relatively consistent when $k_BT/\left(\mathcal{E}_0 S\right)$ is less than 0.16.
As the temperature increases, the total thermal conductivity of the three lowest bands remains negative,
while the thermal conductivity of the total eighteen bands begins to increase and changes its sign at $k_BT/\left(\mathcal{E}_0 S\right) \approx 0.21$,
indicating that higher magnon bands begin to play a vital role afterwards.

\begin{thebibliography}{99}
\bibitem{McClarty2022}
P. A. McClarty,
Topological Magnons: A Review,
\href{https://doi.org/10.1146/annurev-conmatphys-031620-104715}{Annu. Rev. Condens. Matter Phys. \textbf{13}, 171 (2022). }


\bibitem{PengYan2021}
Z.-X. Li, Y. Cao, and P. Yan,
Topological insulators and semimetals in classical magnetic systems,
\href{https://doi.org/10.1016/j.physrep.2021.02.003}{Phys. Rep. \textbf{915}, 8 (2021).}

\bibitem{Fengjun2023}
F. Zhuo, J. Kang, A. Manchon, and Z. Cheng,
Topological phases in magnonics: A review,
\href{https://doi.org/10.1002/apxr.202300054}{Adv. Phys. Res. \textbf{2023}, 2300054 (2023).}
	
\bibitem{Lenk2011}
B. Lenk, H. Ulrichs, F. Garbs, and M. M\"unzenberg,
The building blocks of magnonics,
\href{https://doi.org/10.1016/j.physrep.2011.06.003}{Phys. Rep. \textbf{507}, 107 (2011).}

\bibitem{Chumak2015}
A. V. Chumak, V. I. Vasyuchka, A. A. Serga, and B. Hillebrands,
Magnon spintronics,
\href{https://doi.org/10.1038/nphys3347}{Nat. Phys. \textbf{11}, 453 (2015).}
	
\bibitem{Lifa2013}
L. Zhang, J. Ren, J.-S. Wang, and B. Li,
Topological magnon insulator in insulating ferromagnet,
\href{https://doi.org/10.1103/PhysRevB.87.144101}{Phys. Rev. B. \textbf{87}, 144101 (2013).}

\bibitem{Shindou2013}
R. Shindou, R. Matsumoto, S. Murakami, and J.-i. Ohe,
Topological chiral magnonic edge mode in a magnonic crystal,
\href{https://doi.org/10.1103/PhysRevB.87.174427}{Phys. Rev. B. \textbf{87}, 174427 (2013).}

\bibitem{Mook2014}
A. Mook, J. Henk, and I. Mertig,
Edge states in topological magnon insulators,
\href{https://doi.org/10.1103/PhysRevB.90.024412}{Phys. Rev. B. \textbf{90}, 024412 (2014).}

\bibitem{Nakata2017}
K. Nakata, S. K. Kim, J. Klinovaja, and D. Loss,
Magnonic topological insulators in antiferromagnets,
\href{https://doi.org/10.1103/PhysRevB.96.224414}{Phys. Rev. B. \textbf{96}, 224414 (2017).}

\bibitem{Katsura2010}
H. Katsura, N. Nagaosa, and P. A. Lee,
Theory of the Thermal Hall Effect in Quantum Magnets,
\href{https://doi.org/10.1103/PhysRevLett.104.066403}{Phys. Rev. Lett. \textbf{104}, 066403 (2010).}


\bibitem{Qin2011}
T. Qin, Q. Niu, and J. Shi,
Energy Magnetization and the Thermal Hall Effect,
\href{https://doi.org/10.1103/PhysRevLett.107.236601}{Phys. Rev. Lett. \textbf{107}, 236601 (2011).}

\bibitem{Matsumoto2011prl}
R. Matsumoto and S. Murakami,
Theoretical Prediction of a Rotating Magnon Wave Packet in Ferromagnets,
\href{https://doi.org/10.1103/PhysRevLett.106.197202}{Phys. Rev. Lett. \textbf{106}, 197202 (2011).}

\bibitem{Ideue2012}
T. Ideue, Y. Onose, H. Katsura, Y. Shiomi, S. Ishiwata, N. Nagaosa, and Y. Tokura,
Effect of lattice geometry on magnon Hall effect in ferromagnetic insulators,
\href{https://doi.org/10.1103/PhysRevB.85.134411}{Phys. Rev. B. \textbf{85}, 134411 (2012).}

\bibitem{Matsumoto2014}
R. Matsumoto, R. Shindou, and S. Murakami,
Thermal Hall effect of magnons in magnets with dipolar interaction,
\href{https://doi.org/10.1103/PhysRevB.89.054420}{Phys. Rev. B. \textbf{89}, 054420 (2014).}

\bibitem{Mook2014b}
A. Mook, J. Henk, and I. Mertig,
Magnon Hall effect and topology in kagome lattices: A theoretical investigation,
\href{https://doi.org/10.1103/PhysRevB.89.134409}{Phys. Rev. B. \textbf{89}, 134409 (2014).}

\bibitem{Murakami2017}
S. Murakami and A. Okamoto,
Thermal Hall Effect of Magnons,
\href{https://doi.org/10.7566/JPSJ.86.011010}{J. Phys. Soc. Jpn. \textbf{86}, 011010 (2017).}

\bibitem{Mook2019}
A. Mook, J. Henk, and I. Mertig,
Thermal Hall effect in noncollinear coplanar insulating antiferromagnets,
\href{https://doi.org/10.1103/PhysRevB.99.014427}{Phys. Rev. B \textbf{99}, 014427 (2019).}

\bibitem{Yang2020}
Y.-f. Yang, G.-M. Zhang, and F.-C. Zhang,
Universal Behavior of the Thermal Hall Conductivity,
\href{https://doi.org/10.1103/PhysRevLett.124.186602}{Phys. Rev.  Lett. \textbf{124}, 186602 (2020).}

\bibitem{Wen2020}
H. Xu, S.-f. Cheng, S. Bao, J.-S. Wen,
Experimental Progress in Thermal Hall Conductivity Research on Strongly Correlated Electronic Systems,
\href{https://doi.org/10.13725/j.cnki.pip.2022.05.001}{Progress in Physics \textbf{42}, 159-183 (2022).}

\bibitem{Mook2022}
R. R. Neumann, A. Mook, J. Henk, and I. Mertig,
Thermal Hall Effect of Magnons in Collinear Antiferromagnetic Insulators: Signatures of Magnetic and Topological Phase Transitions,
\href{https://doi.org/10.1103/PhysRevLett.128.117201}{Phys. Rev.  Lett. \textbf{128}, 117201 (2022).}

\bibitem{Yao2023}
M.-H. Zhang, and D.-X. Yao,
Topological magnons on the triangular kagome lattice,
\href{https://doi.org/10.1103/PhysRevB.107.024408}{Phys. Rev. B. \textbf{107}, 024408 (2023).}



\bibitem{Chen2023}
X.-T. Zhang, Y. H. Gao, G. Chen,
Thermal Hall effects in quantum magnets,
\href{https://doi.org/10.1016/j.physrep.2024.03.004}{Phys. Rep. \textbf{1070}, 1 (2024).}


\bibitem{Onose2010}
Y. Onose, T. Ideue, H. Katsura, Y. Shiomi, N. Nagaosa, and Y.
Tokura,
Observation of the Magnon Hall Effect,
\href{https://doi.org/10.1126/science.1188260}{Science \textbf{329}, 297 (2010).}

\bibitem{Chisnell2015}
R. Chisnell, J.S. Helton, D.E. Freedman, D.K. Singh, R.I. Bewley, D.G. Nocera, and Y.S. Lee,
Topological Magnon Bands in a Kagome Lattice Ferromagnet,
\href{https://doi.org/10.1103/PhysRevLett.115.147201}{Phys. Rev.  Lett. \textbf{115}, 147201 (2015).}

\bibitem{Hirschberger2015}
M. Hirschberger, J. W. Krizan, R. J. Cava, and N. P. Ong,
Large thermal Hall conductivity of neutral spin excitations in a frustrated quantum magnet,
\href{https://doi.org/10.1126/science.1257340}{Science \textbf{348}, 106 (2015)}


\bibitem{Hirschberger2015prl}
M. Hirschberger, R. Chisnell, Y. S. Lee, and N.P. Ong,
Thermal Hall Effect of Spin Excitations in a Kagome Magnet,
\href{https://doi.org/10.1103/PhysRevLett.115.106603}{Phys. Rev.  Lett. \textbf{115}, 106603 (2015).}

\bibitem{Trebst2022}
S. Trebst, C. Hickey,
Kitaev materials,
\href{https://doi.org/10.1016/j.physrep.2021.11.003}{Phys. Rep. \textbf{950}, 1 (2022).}	

\bibitem{Winter2017}
S. M Winter, A. A Tsirlin, M. Daghofer, J. van den Brink, Y. Singh, P. Gegenwart and R. Valent\'i,
Models and materials for generalized Kitaev magnetism,
\href{https://doi.org/10.1088/1361-648X/aa8cf5}{J. Phys.: Condens. Matt.  \textbf{29}, 493002 (2017).}

\bibitem{Wen2017}
K. Ran, J. Wang, W. Wang, Z.-Y. Dong, X. Ren, S. Bao, S.
Li, Z. Ma, Y. Gan, Y. Zhang, J. T. Park, G. Deng, S. Danilkin,
S.-L. Yu, J.-X. Li, and J. Wen,
Spin-Wave Excitations Evidencing the Kitaev Interaction in Single Crystalline $\alpha$-RuCl$_3$,
\href{https://doi.org/10.1103/PhysRevLett.118.107203}{Phys. Rev. Lett. \textbf{118}, 107203 (2017).}

\bibitem{Janssen2019}
L. Janssen, and M. Vojta,
Heisenberg$-$Kitaev physics in magnetic fields,
\href{https://doi.org/10.1088/1361-648X/ab283e}{J. Phys.: Condens. Matt.  \textbf{31}, 423002 (2019).}

\bibitem{XuFXetal2018}
C. Xu, J. Feng, H. Xiang, and L. Bellaiche,
Interplay between Kitaev interaction and single ion anisotropy in ferromagnetic $\rm CrI_3$ and $\rm CrGeTe_3$ monolayers,
\href{https://doi.org/10.1038/s41524-018-0115-6}{npj Comput. Mater. \textbf{4}, 57 (2018).}

\bibitem{LeeUWetal2020}
I. Lee, F. G. Utermohlen, D. Weber, K. Hwang, C. Zhang, J. van Tol, J. E. Goldberger, N. Trivedi, and P. C. Hammel,
Fundamental spin interactions underlying the magnetic anisotropy in the Kitaev ferromagnet $\rm CrI_3$,
\href{https://doi.org/10.1103/PhysRevLett.124.017201}{Phys. Rev. Lett. \textbf{124}, 017201 (2020).}

\bibitem{GLin2021}
G. Lin, \textit{et al.},
Field-induced quantum spin disordered state in spin-1/2 honeycomb magnet Na$_2$Co$_2$TeO$_6$,
\href{https://doi.org/10.1038/s41467-021-25567-7}{Nat. Commun. \textbf{12}, 5559 (2021).}

\bibitem{Li2022}
  X. Li, Y. Gu, Y. Chen, V. O. Garlea, K. Iida, K. Kamazawa, Y. Li, G. Deng, Q. Xiao, X. Zheng, Z. Ye, Y. Peng, I. A. Zaliznyak, J. M. Tranquada, and Y. Li,
  Giant Magnetic In-Plane Anisotropy and Competing Instabilities in Na$_3$Co$_2$SbO$_6$,
  \href{https://doi.org/10.1103/PhysRevX.12.041024}{Phys. Rev. X \textbf{12}, 041024 (2022).}

\bibitem{Yao2023prr}
  W. Yao, Y. Zhao, Y. Qiu, C. Balz, J. R. Stewart, J. W. Lynn, and Y. Li,
  Magnetic ground state of the Na$_2$Co$_2$TeO$_6$ Kitaev spin liquid candidate,
  \href{https://doi.org/10.1103/PhysRevResearch.5.L022045}{Phys. Rev. Res. \textbf{5}, L022045 (2023).}

\bibitem{Jie2023}
G. Lin, J. Jiao, X. Li, M. Shu, O. Zaharko, T. Shiroka, T. Hong, A. I. Kolesnikov, G. Deng, S. Dunsiger, H. Zhou, T. Shang, and J. Ma,
Static magnetic order with strong quantum fluctuations in spin-1/2 honeycomb magnet Na$_2$Co$_2$TeO$_6$,
\href{https://arxiv.org/abs/2312.06284}{arXiv:2312.06284 (2023).}


\bibitem{WangNmdLiu2019}
  J. Wang, B. Normand, and Z.-X. Liu,
  One Proximate Kitaev Spin Liquid in the $K$-$J$-$\Gamma$ Model on the Honeycomb Lattice,
  \href{https://doi.org/10.1103/PhysRevLett.123.197201}{Phys. Rev. Lett. \textbf{123}, 197201 (2019).}

\bibitem{RalkoMerino2020}
  A. Ralko and J. Merino,
  Novel Chiral Quantum Spin Liquids in Kitaev Magnets,
  \href{https://doi.org/10.1103/PhysRevLett.124.217203}{Phys. Rev. Lett. \textbf{124}, 217203 (2020).}


\bibitem{LuoNPJ2021}
  Q. Luo, J. Zhao, H.-Y. Kee, and X. Wang,
  Gapless quantum spin liquid in a honeycomb $\Gamma$ magnet,
  \href{https://doi.org/10.1038/s41535-021-00356-z}{npj Quantum Mater. \textbf{6}, 57 (2021).}

\bibitem{LeeKCetal2020}
  H.-Y. Lee, R. Kaneko, L. E. Chern, T. Okubo, Y. Yamaji, N. Kawashima, and Y. B. Kim,
  Magnetic-field induced quantum phases in tensor network study of Kitaev magnets,
  \href{https://doi.org/10.1038/s41467-020-15320-x}{Nat. Commun. \textbf{11}, 1639 (2020).}

\bibitem{GohlkeCKK2020}
  M. Gohlke, L.~E. Chern, H.-Y. Kee, and Y.~B. Kim,
  Emergence of nematic paramagnet via quantum order-by-disorder and pseudo-Goldstone modes in Kitaev magnets,
  \href{https://doi.org/10.1103/PhysRevResearch.2.043023}{Phys. Rev. Research \textbf{2}, 043023 (2020).}

\bibitem{McClarty2018}
P. A. McClarty, X.-Y. Dong, M. Gohlke, J. G. Rau, F. Pollmann, R. Moessner, and K. Penc,
Topological magnons in Kitaev magnets at high fields,
\href{https://doi.org/10.1103/PhysRevB.98.060404}{Phys. Rev. B. \textbf{98}, 060404(R) (2018).}

\bibitem{Joshi2018}
D. G. Joshi,
Topological excitations in the ferromagnetic Kitaev-Heisenberg model,
\href{https://doi.org/10.1103/PhysRevB.98.060405}{Phys. Rev. B. \textbf{98}, 060405(R) (2018).}

\bibitem{Luo2020SciPost}
Z.-X. Luo and G. Chen,
  Honeycomb rare-earth magnets with anisotropic exchange interactions
  \href{https://scipost.org/SciPostPhysCore.3.1.004}{SciPost Phys. Core \textbf{3}, 004 (2020).}


\bibitem{Chern2020}
L. E. Chern, R. Kaneko, H.-Y. Lee, and Y. B. Kim,
Magnetic field induced competing phases in spin-orbital entangled Kitaev magnets,
\href{https://doi.org/10.1103/PhysRevResearch.2.013014}{Phys. Rev. Research \textbf{2}, 013014 (2020).}

\bibitem{Aguilera2020}
E. Aguilera, R. Jaeschke-Ubiergo, N. Vidal-Silva, Luis E. F. Foa Torres, and A. S. Nunez,
Topological magnonics in the two-dimensional van der Waals magnet  CrI$_3$,
\href{https://doi.org/10.1103/PhysRevB.102.024409}{Phys. Rev. B \textbf{102}, 024409 (2020).}

\bibitem{Emily2020}
E. Z. Zhang, L. E. Chern, and Y. B. Kim,
Topological magnons for thermal Hall transport in frustrated magnets with bond-dependent interactions,
\href{https://doi.org/10.1103/PhysRevB.103.174402}{Phys. Rev. B \textbf{103}, 174402 (2021).}


\bibitem{Chern2021prl}
L. E. Chern, E. Z. Zhang, and Y. B. Kim
Sign Structure of Thermal Hall Conductivity and Topological Magnons for In-Plane Field Polarized Kitaev Magnets,
\href{https://doi.org/10.1103/PhysRevLett.126.147201}{Phys. Rev. Lett. \textbf{126}, 147201 (2021).}


\bibitem{Kasahara2018Nature}
Y. Kasahara, T. Ohnishi, Y. Mizukami, O. Tanaka, Sixiao Ma, K. Sugii, N. Kurita, H. Tanaka, J. Nasu, Y. Motome, T. Shibauchi, and Y. Matsuda,
  Majorana quantization and half-integer thermal quantum Hall effect in a Kitaev spin liquid,
  \href{https://www.nature.com/articles/s41586-018-0274-0}{Nature \textbf{559}, 227 (2018).}

\bibitem{Czajka2021NatPhys}
P. Czajka, T. Gao, M. Hirschberger, P. Lampen-Kelley, A. Banerjee, J. Yan, D. G. Mandrus, S. E. Nagler, and N. P. Ong,
  Oscillations of the thermal conductivity in the spin-liquid state of $\alpha$- RuCl$_3$,
  \href{https://www.nature.com/articles/s41567-021-01243-x}{Nat. Phys. \textbf{11}, 915 (2021).}

\bibitem{Kee2023NatMater}
H.-Y. Kee,
  Thermal Hall conductivity of $\alpha$- RuCl$_3$,
  \href{https://www.nature.com/articles/s41563-022-01444-6}{Nat. Mater. \textbf{22}, 6 (2023).}


\bibitem{Takeda2022PRR}
H. Takeda, J. Mai, M. Akazawa, K. Tamura, J. Yan, K. Moovendaran, K. Raju, R. Sankar, K.-Y. Choi, and M. Yamashita,
  Planar thermal Hall effects in the Kitaev spin liquid candidate Na$_2$Co$_2$TeO$_6$,
  \href{https://doi.org/10.1103/PhysRevResearch.4.L042035}{Phys. Rev. Research \textbf{4}, L042035 (2022).}

\bibitem{Guang2023PRB}
S. Guang, N. Li, R. L. Luo, Q. Huang, Y. Wang, X. Yue, K. Xia, Q. Li, X. Zhao, G. Chen, H. Zhou, and X. Sun,
  Thermal transport of fractionalized antiferromagnetic and field-induced states in the Kitaev material Na$_2$Co$_2$TeO$_6$,
  \href{https://doi.org/10.1103/PhysRevB.107.184423}{Phys. Rev. B \textbf{107}, 184423 (2023).}

\bibitem{Luo2022PRB}
  Q. Luo and H.-Y. Kee,
  Interplay of magnetic field and trigonal distortion in the honeycomb $\Gamma$ model: Occurrence of a spin-flop phase,
  \href{https://doi.org/10.1103/PhysRevB.105.174435}{Phys. Rev. B \textbf{105}, 174435 (2022).}


\bibitem{Rau2014}
  J. G. Rau, E. K.-H. Lee, and H.-Y. Kee,
  Generic Spin Model for the Honeycomb Iridates beyond the Kitaev Limit,
  \href{https://doi.org/10.1103/PhysRevLett.112.077204}{Phys. Rev.  Lett. \textbf{112}, 077204 (2014).}


\bibitem{Chern2021npj}
L. E. Chern, Finn L. Buessen, and Y. B. Kim,
Classical magnetic vortex liquid and large thermal Hall conductivity in frustrated magnets with bond-dependent interactions,
\href{https://doi.org/10.1038/s41535-021-00331-8}{npj Quantum Mater. \textbf{6}, 33 (2021).}


\bibitem{Liu2021PRR}
  K. Liu, N. Sadoune, Nihal Rao, J. Greitemann, and L. Pollet,
  Revealing the phase diagram of Kitaev materials by machine learning: Cooperation and competition between spin liquids,
  \href{https://doi.org/10.1103/PhysRevResearch.3.033223}{Phys. Rev.  Research \textbf{3}, 023016 (2021).}

\bibitem{Rao2021PRR}
  N. Rao, K. Liu, M. Machaczek, and L. Pollet,
  Machine-learned phase diagrams of generalized Kitaev honeycomb magnets,
  \href{https://doi.org/10.1103/PhysRevResearch.3.033223}{Phys. Rev.  Research \textbf{3}, 033223 (2021).}


\bibitem{Rayyan2022}
A. Rayyan, Q. Luo, and H.-Y. Kee,
Extent of frustration in the classical Kitaev-$\Gamma$ model via bond anisotropy,
\href{https://doi.org/10.1103/PhysRevB.104.094431}{Phys. Rev. B \textbf{104}, 094431 (2021).}

\bibitem{Ken2023}
K. Chen, Q. Luo, Z. Zhou S. He, B. Xi, C. Jia, H.-G. Luo and J. Zhao,
Triple-meron crystal in high-spin Kitaev magnets,
\href{https://doi.org/10.1088/1367-2630/acb5bb}{New J. Phys. \textbf{25}, 023006 (2023).}

\bibitem{Stavropoulos2023arXiv}
  P.~P. Stavropoulos, Y. Yang, I. Rousochatzakis, and N. B. Perkins,
  Complex orders and chirality in the classical Kitaev-$\Gamma$ model,
  \href{https://arxiv.org/abs/arXiv:2206.08946}{arXiv:2311.00037 (2023).}

\bibitem{Rousochatzakis2023}
  I. Rousochatzakis, N. B. Perkins, Q. Luo, and H. Y. Kee,
  Beyond Kitaev physics in strong spin-orbit coupled magnets,
  \href{https://doi.org/10.1088/1361-6633/ad208d}{Rep. Prog. Phys. \textbf{87}, 026502 (2024).}

\bibitem{HukushimaNemoto1996}
K. Hukushima and K. Nemoto,
Exchange Monte Carlo method and application to spin glass simulations,
\href{https://doi.org/10.1143/JPSJ.65.1604}{J. Phys. Soc. Jpn. \textbf{65}, 1604 (1996).}


\bibitem{Janssen2016}
L. Janssen, E. C. Andrade, and M. Vojta,
Honeycomb-Lattice Heisenberg-Kitaev Model in a Magnetic Field: Spin Canting, Metamagnetism, and Vortex Crystals,
\href{https://doi.org/10.1103/PhysRevLett.117.277202}{Phys. Rev. Lett. \textbf{117}, 277202 (2016).}

\bibitem{Toth2015}
S. Toth and B. Lake,
Linear spin wave theory for single-Q incommensurate magnetic structures,
\href{https://doi.org/10.1088/0953-8984/27/16/166002}{J. Phys.: Condens. Matter \textbf{27}, 166002 (2015).}

\bibitem{Daniel2019}
S. A. D\'iaz, J. Klinovaja, and D. Loss,
Topological Magnons and Edge States in Antiferromagnetic Skyrmion Crystals,
\href{https://doi.org/10.1103/PhysRevLett.122.187203}{Phys. Rev.  Lett. \textbf{122}, 187203 (2019).}


\bibitem{Takuya2020}
T. Matsumoto, and S. Hayami,
Nonreciprocal magnons due to symmetric anisotropic exchange interaction in honeycomb antiferromagnets,
\href{https://doi.org/10.1103/PhysRevB.101.224419}{Phys. Rev. B. \textbf{101}, 224419 (2020).}

\bibitem{Satoru2022}
S. Hayami, and T. Matsumoto,
Essential model parameters for nonreciprocal magnons in multisublattice systems,
\href{https://doi.org/10.1103/PhysRevB.105.014404}{Phys. Rev. B. \textbf{105}, 014404 (2022).}

\bibitem{Hatsugai1993}
Y. Hatsugai,
Chern number and edge states in the integer quantum Hall effect,
\href{https://doi.org/10.1103/PhysRevLett.71.3697}{Phys. Rev.  Lett. \textbf{71}, 3697 (1993).}

\bibitem{Chernyshev2013}
M. E. Zhitomirsky and A. L. Chernyshev,
Colloquium: Spontaneous magnon decays,
\href{https://doi.org/10.1103/RevModPhys.85.219}{Rev. Mod. Phys. \textbf{85}, 219 (2013).}

\bibitem{Mook2020prr}
A. Mook, J. Klinovaja, and D. Loss,
Quantum damping of skyrmion crystal eigenmodes due to spontaneous quasiparticle decay,
\href{https://doi.org/10.1103/PhysRevResearch.2.033491}{Phys. Rev. Research \textbf{2}, 033491 (2020).}

\bibitem{Mook2021prx}
A. Mook, K. Plekhanov, J. Klinovaja, and D. Loss,
Interaction-Stabilized Topological Magnon Insulator in Ferromagnets,
\href{https://doi.org/10.1103/PhysRevX.11.021061}{Phys. Rev. X \textbf{11}, 021061 (2021).}

\bibitem{Habel2024}
J. Habel, A. Mook, J. Willsher, and J. Knolle,
Breakdown of chiral edge modes in topological magnon insulators,
\href{https://doi.org/10.1103/PhysRevB.109.024441}{Phys. Rev. B \textbf{109}, 024441 (2024).}

\bibitem{Lu2021}
Y.-S. Lu, J.-L. Li, and C.-T. Wu,
Topological Phase Transitions of Dirac Magnons in Honeycomb Ferromagnets,
\href{https://doi.org/10.1103/PhysRevLett.127.217202}{Phys. Rev.  Lett. \textbf{127}, 217202 (2021).}

\bibitem{Koyama2023}
S. Koyama and J. Nasu,
Flavor-wave theory with quasiparticle damping at finite temperatures: Application to chiral edge modes in the Kitaev model,
\href{https://doi.org/10.1103/PhysRevB.108.235162}{Phys. Rev. B \textbf{108}, 235162 (2023).}

\bibitem{Verresen2019}
R. Verresen,R. Moessner, F. Pollmann,
Avoided quasiparticle decay from strong quantum interactions,
\href{https://doi.org/10.1038/s41567-019-0535-3}{Nat. Phys. \textbf{15}, 750 (2019).}

\bibitem{Gohlke2023}
M. Gohlke, A. Corticelli, R. Moessner, P. A. McClarty, and A. Mook,
Spurious Symmetry Enhancement in Linear Spin Wave Theory and Interaction-Induced Topology in Magnons,
\href{https://doi.org/10.1103/PhysRevLett.131.186702}{Phys. Rev.  Lett. \textbf{131}, 186702 (2023).}

\bibitem{Sun2023}
N. Li, R. R. Neumann, S. K. Guang, Q. Huang, J. Liu, K. Xia, X. Y. Yue, Y. Sun, Y. Y. Wang, Q. J. Li, Y. Jiang, J. Fang, Z. Jiang, X. Zhao, A. Mook, J. Henk, I. Mertig, H. D. Zhou, and X. F. Sun,
Magnon-polaron driven thermal Hall effect in a Heisenberg-Kitaev antiferromagnet,
\href{https://doi.org/10.1103/PhysRevB.108.L140402}{Phys. Rev. B \textbf{108}, L140402 (2023).}

\bibitem{Park2023}
Y. Choi, H. Yang, J. Park, and J.-G. Park,
Sizable suppression of magnon Hall effect by magnon damping in Cr$_2$Ge$_2$Te$_6$,
\href{https://doi.org/10.1103/PhysRevB.107.184434}{Phys. Rev. B \textbf{107}, 184434 (2023).}

\bibitem{Ke2024}
C. Xu, H. Zhang, C. Carnahan, P. Zhang, D. Xiao, and X. Ke,
Thermal Hall effect in the van der Waals ferromagnet CrI$_3$,
\href{https://doi.org/10.1103/PhysRevB.109.094415}{Phys. Rev. B \textbf{109}, 094415 (2024).}

\end{thebibliography}
\end{document}